\documentclass[twocolumn,showpacs,pre, floatfix]{revtex4}

\usepackage{graphicx}
\usepackage{amsmath}
\usepackage[center]{subfigure}

\begin{document}

\title{Fixed points and limit cycles in the population dynamics of lysogenic viruses and their hosts}
\author{Zhenyu Wang and Nigel Goldenfeld}
\affiliation{Department of Physics, University of Illinois at Urbana-Champaign, Loomis
Laboratory of Physics, 1110 West Green Street, Urbana, Illinois, 61801 and\\
Institute for Genomic Biology, University of Illinois at Urbana-Champaign,
1206 West Gregory Drive, Urbana, IL 61801}

\begin{abstract}
Starting with stochastic rate equations for the fundamental
interactions between microbes and their viruses, we derive a mean-field
theory for the population dynamics of microbe-virus systems, including
the effects of lysogeny.  In the absence of lysogeny, our model is a
generalization of that proposed phenomenologically by Weitz and Dushoff.  In
the presence of lysogeny, we analyze the possible states of the system,
identifying a novel limit cycle, which we interpret physically.  To
test the robustness of our mean field calculations to demographic
fluctuations, we have compared our results with stochastic simulations
using the Gillespie algorithm. Finally, we estimate the range of
parameters that delineate the various steady states of our model.
\end{abstract}

\pacs{87.23.Cc, 87.18.Tt, 87.18.Nq}

%
%
%
%
%
%
%
%
%
%
%
%
%

\maketitle

\section{Introduction}

Microbes and their viruses are the most genetically diverse, abundant
and widely distributed organisms across the
planet\cite{MARA95,CHIB04,SANT08,ORTM05}. Microbes are major
contributors to the global biogeochemical cycles and catalyze the
reactions that have over evolutionary time brought the Earth's surface
to its present redox state\cite{FALK08}.  Similarly, viruses, especially
in the oceans, manipulate marine communities through predation and
horizontal gene transfer\cite{syvanen1994hgt,ochman2000lgt}, recycle
nutrients and thus drive the biological pump which leads {\it inter
alia\/} to the sequestration of carbon in the deep
ocean\cite{WILH99,SUTT05,weinbauer2004vdm,SHAR06,MONS06,SCHI07,SUTT07,DANO08,BARD08,ROHW09}.

It is being increasingly realized that the classical view of microbial
viruses purely as predators is too limited.  Many microbe-virus
interactions are lysogenic, not lytic: upon infection, the viral
genetic material is incorporated into the chromosome of the host,
replicates with the host, and can be subsequently released, typically
triggered by the stress response of the host to environmental
change\cite{PAUL08}. As a result, viruses can transfer genes to and
from bacteria, as well as being predators of them, so that the
virosphere should properly be recognized as a massive gene
reservoir\cite{sonea1988bwl,sullivan2006pae,GOLD07,ROHW09}.  Thus there
is a coevolution of both communities, the effects of which are complex
and
far-reaching\cite{ANDE66,ANDE70,filee2003rpv,weinbauer2004vdm,sullivan2006pae,LIND07,GOLD07,ROHW09},
even including the manipulation of bacterial mutation
rates\cite{PAL07}. This nontrivial interaction between microbes and
viruses has not gone unnoticed, with wide interest among biologists,
ecologists and geologists\cite{HADA97,WILL02,BRUS04,WILL04,
CHIB04,BRUS05,ROSV06,LIND07,PAUL08,WILL08,WEIT08}.

These findings highlight the importance of considering ecosystem
dynamics within an evolutionary context.  Conversely, evolution needs
to be properly understood as arising from a spatially-resolved
ecological context, as was first recognized by Wallace over 150 years
ago\cite{wallace1855law}.  That speciation, and adaptation in general,
arises at a particular point in time and space has a number of deep
consequences that have not yet been incorporated into current theory.
First, it means that evolutionary dynamics proceeds by the propagation
of fronts, resulting in a complex and dynamical pattern of speciation,
adaptation and genome divergence that reflects its intrinsic dynamics
and that of the heterogeneous and dynamical
environment\cite{ibrahim1996spatial,hallatschek2008gene,hallatschek2009fisher,korolev2009}.
Second, as fronts expand, there are only a few pioneer organisms at the
leading edge, and so demographic fluctuations are much larger than in
the bulk.  Such fluctuations profoundly influence the spatial structure
of the populations, and during the last few years have been recognized
to play a major role in population cycles\cite{MCKA04} and even spatial
pattern formation\cite{butler2009robust}. Third, the existence of
horizontal gene transfer and genome rearrangement processes is strongly
coupled to spatial distribution. For example, it is known that the
probability of conjugation events is dependent on the local density,
being essentially one/per generation in closely-packed biofilms, but an
order of magnitude smaller in planktonic culture\cite{babic2008direct}.
Moreover, the mechanism of horizontal gene transfer is also dependent
on the density, with viral-mediated transduction being the most
relevant mechanism at low density.  How these patterns of evolutionary
dynamics and species distribution play out is essentially unexplored.
However, there have recently been the first reports of observations of
the coupling between evolutionary and ecological timescales.  In one
such system (a predator-prey system realized in rotifer-algae
interactions), it has been demonstrated that rapid evolutionary
dynamics is responsible for the unusual phase-lag characteristics of
the observed population oscillations\cite{yoshida2007cryptic}. Thus,
rapid evolution is not only a major force for adaptation, but can have
marked ecological consequences too.

The complex interplay between evolution and the environment is nowhere
more important than in early life, where the key questions concern how
life emerged from abiotic geochemistry.  Early life experienced
demanding environments, whose closest modern day correspondence might
be deep ocean hydrothermal vents or hot springs. It is known that there
are high occurrence of lysogens in both
environments\cite{ORTM05,held2009viral}, suggesting that microbe-phage
interactions might also be important in the early stages of life, with
lysogens playing an important role as a reservoir of genes and perhaps
even aiding in the stabilitization of early life populations through
the limit cycle mechanism discussed in this paper.

Our goal in this paper is to lay a theoretical foundation for
describing the interplay between ecology and evolution in the context
of microbe-virus systems, as these are arguably amongst the most
important and probably the simplest of the complex systems in biology.
The questions that will ultimately interest us are the evolutionary
pressures that tune genetic switches governing the lysis-lysogeny
decision, as well as the factors that shape prophage induction in
response to environmental
stress\cite{lenski1994evolution,koch2007evolution,refardt2009tuning}.
Such a foundation must begin with a proper account of the population
dynamics itself, before coupling in detail to other levels of
description involving genome dynamics, for example.  Thus, we have
chosen to focus in the present paper on the dynamics of microbe-virus
systems, taking full account of both of the major viral pathways.  In
this paper, we are not specific about whether we are dealing with
bacterial or archaeal viruses, but because most of the experiments to
date are carried out on bacteria, we have tended to identify the
microbes as bacteria and the viruses as phages, even though this is not
required by the mathematics.

We are now ready to introduce the specific problem that we treat in
this paper. Upon phage infection, there are two pathways awaiting the
host bacterium\cite{PTAS04}. In the first pathway---lysis---the
bacteriophage produces a large number of copies of itself utilizing the
bacterium's genetic material and molecular machinery. As a result, the
bacterium ceases its metabolic function, and ruptures, releasing the
newly-assembled bacteriophage inside.  The other pathway is lysogeny.
In this process, the intruder integrates its own DNA into the genome of
the bacterium, enters a dormant stage and becomes a prophage. The
infected bacterium is known as a lysogen---a relatively stable
state\cite{BAEK03}, immune to superinfection from the same or related
strains, and proceeding under normal replication life-cycles. The DNA
of the bacteriophage is duplicated, along with that of the host during
cell replication. The lysogenic state can be terminated by
environmental stress such as starvation, pollution or ultraviolet
irradiation, resulting in the process known as prophage induction: the
exit of the prophage from the host genome, and subsequent lysis of the
original bacterium and its bacterial descendants.

We now discuss briefly existing treatments of population dynamics in
the context of microbe-virus systems.  In 1977, Levin \emph{et al}
\cite{LEVI77} extended the celebrated Lotka-Volterra equations to model
the dynamics between virulent phages and their victims, where only
virulent phages are considered. A number of extensions have been
proposed, extending the level of biological realism to include such
features as the time delay arising between infection and lysis as well
as the evolution of kinetic
parameters\cite{WANG96,BOHA97,BERE01,FORT02,WEIT05}. In 2007, Weitz and
Dushoff\cite{WEIT08b} proposed another way to improve the classic
predator-prey model. Their attempt was mainly based on the experimental
observation that the ability of a bacteriophage to lyse hosts degrades
when the bacteria approach their carrying
capacity\cite{MIDD00,BRUS04a,SILL04}. Adding a new term to account for
the saturation of the infection of the bacteriophages, they obtained an
interesting phase diagram in which the fate of the
bacteria-phage community can depend on the initial conditions. However,
the new term is put in by hand, based on intuition which needs detailed
mathematical support. Furthermore, they focused on virulent phages and
excluded the temperate ones that elicit lysogeny, now regarded as
essential to the survival of microbial communities through fluctuating
environments\cite{BRUS04,PAUL08, WILL08}.

These works are based upon an ensemble-level description of the
community, as in the classic work on predator-prey
systems\cite{smith1978models}.  However, as is well-known
\cite{smith1978models}, the simplest of these models fails to capture
the intrinsic cyclical behavior of predator-prey populations despite
apparently incorporating fully the basic interactions that should give
rise to cycles.  This paradox was resolved by the important work of
McKane and Newman\cite{MCKA04}, who showed that cyclical effects could
only be captured at the level of an individual-level model, and arose
through the amplification of demographic noise.  Their work showed how
the conventional ensemble-level equations for predator-prey systems
arose as the mean field limit of the appropriate statistical field
theory, with the essential effects of demographic noise entering the
analysis as one-loop corrections to mean field theory, in an inverse
population size expansion.  These effects can also be treated in a
slightly more convenient formalism using path
integrals\cite{butler2009predator}.  The literature also does not have
an explicit representation of lysogeny as it modifies the population
dynamics of both host and phage.

The purpose of this paper is to provide a detailed theory of the
population dynamics for host-phage communities.  In contrast to earlier
work, we pose the problem at the microscopic level, working with an
individual-level model of bacteria and phage.  From this fundamental
description, we are able to derive the usual community level
description analogous to Lotka-Volterra equations from a mean field
theory.  Our results encompass both virulent phages, such as those in
Weitz and Dushoff's work\cite{WEIT08b}, and lysogenic phages which have
not been studied mathematically up to now.

This paper is organized as follows. In Section \ref{lysis-only}, as a
preliminary exercise, to present the technique, we treat a lysis-only
model, in which we derive a set of dynamical equations roughly in the
same form as in Weitz and Dushoff's paper\cite{WEIT08b} except for an
additional parameter, which generally results in a relatively
unimportant shift in the phase diagram. In the full lysogeny-lysis
model, presented in Section \ref{lysogeny-lysis}, we develop the
formalism for the community of hosts and phages, including both lysis
and lysogeny. Interestingly, we find that for certain combination of
parameters, the community exhibits a limit cycle for all the species in
the phase space, even at the level of mean field theory. In order to
interpret the corresponding range of parameters in a useful way for
experimental observations, we map the parameters to rates in chemical
reactions.  In order to explore the robustness of our results, we
demonstrate in Section \ref{stochastic-simulation} that the
corresponding limit cycle arises also in stochastic simulations with
the Gillespie algorithm. Finally, in Section \ref{parameters} we
estimate the feasibility of verifying our predictions in laboratory
experiments.

\section{Lysis-only model}
\label{lysis-only}

\subsection{Derivation of the population dynamics from an individual-level model}

In this section, we adapt the classic predator-prey model to the
host-phage communities from a microscopic or individual-level model.
For simplicity, we first focus on two-component competition,
where lysogens are excluded in spite of their
biological importance. Hence, we are considering virulent phages and
their hosts. Following the procedure given by McKane and
Newman\cite{MCKA04}, we derive
the population dynamics for the host-phage system, which Weitz and Dushoff%
\cite{WEIT08b} had written down based on intuition. Here we work at the
level of mean field theory, and we do not, in this paper, include the
extension necessary for representing spatial degrees of freedom.  In
our model, the host-phage dynamics differentiates itself from the
classic predator-prey model in two ways: (1) only the host population
is restricted by carrying capacity due to resource limitation and (2)
the lysis of one host releases a particular number of phages (for
example, about 100 replicates for lambda phage\cite{PTAS04}), instead of
only one predator in the classic predator-prey model. The above two
points need to be accounted for carefully in the set up of the model,
especially in the introduction and application of the carrying
capacity, which will be explained explicitly as follows.

In our host-phage community, we have only one species of host and one
species of phage which preys upon the former. Let us label the hosts by
A and phages by B, whose populations are $m$ and $n$, respectively. The
hosts, either heterotrophic or autotrophic, need to consume
environmental resources, which are renewable in every cycle, for
survival and reproduction. All the environmental limitations on the
hosts are abstracted into a maximal host population, which is denoted
by the carrying capacity $K$. The phages, on the other hand, do not
rely on the consumption of natural resources for maintenance once they
are released into the environment. Thus, there is no such hard
constraint on the phage population. Although phages are not restricted
by $K$, we still introduce a virtual carrying capacity $W$ for phages
from dimensional considerations. It can be imagined that $W\rightarrow
\infty $ so that no true carrying capacity is imposed on the phage
population. The carrying capacities can be better visualized if we
conceive space to be equally divided into $K$ units for hosts and $W$
units for phages.  These units will be referred to as the host layer
and phage layer, respectively. In the host layer, each unit is either
occupied by one host or unoccupied, i.e. an empty site $E$. The total
number of empty sites $E$ is $K-m$. We construct the phage layer in a
similar manner and denote the empty sites there by $\phi$ although the
phage population is not confined actually.

The population dynamics of the system can thus be modeled as arising from the
following six microscopic events (Table \ref{tab:lysis_events}):

\begin{table}[htbp]
\caption{Microscopic events in the lysis-only model.}
\label{tab:lysis_events}\centering
\begin{tabular}{lc}
\hline\hline
description & symbols \\[0.5ex] \hline
&  \\[-2.6ex]
birth of host & $AE\overset{b}{\rightarrow }AA$ \\[0.5ex]
death of host due to longevity & $A\overset{c}{\rightarrow }E$ \\[0.5ex]
death of host due to crowding effect & $AA\overset{d}{\rightarrow }AE$ \\%
[0.5ex]
host-phage interaction &  \\[0.5ex]
$\cdot$ under good metabolism & $AEB\overset{e}{\rightarrow }EE\alpha B$ \\%
[0.5ex]
$\cdot$ under poor metabolism & $AAB\overset{f}{\rightarrow }EA\beta B$ \\%
[0.5ex]
death of phage & $B\overset{g}{\rightarrow }\phi $ \\[1ex] \hline
\end{tabular}%
\end{table}

Here, $b$, $c$, $d$, $e$, $f$ and $g$ are all constant rates.
All the events above are written with constraints, with a nonlinear
relation being incorporated automatically by adding empty sites $E$ to the
left of the arrows to reflect the restriction of carrying capacity $K$.
For example, the birth of the host is density dependent, which needs an
empty site to accommodate the newly-born host. If no empty site is
found, such an event can not happen. Since we consider only the mean
field case, no spatial inhomogeneity is introduced. There is no concept
of locality here, either. As long as an empty site is found, the
newly-born host is permitted. The crowding effect describes the competition
in survival for limited natural resources among hosts. No such
crowding effect exists for phages, which is in line with our assumption
that there is no true carrying capacity confining the phage
population. The two events in host-phage interaction are carefully
chosen to give a minimal model while encompassing reduced lysis when
the host population is approaching its carrying capacity. In the
events, $\alpha $ and $\beta $ are the numbers of progeny for phage
reproduction under good and poor metabolism, respectively. In biology,
there are mainly two reasons which may account for reduced lysis.
The first is the decrease in the phage reproduction
number\cite{MIDD00}, i.e.
\begin{equation}
\alpha >\beta ,  \label{ab}
\end{equation}
because phages need bacterial genetic materials, molecular machinery
and energy in the synthesis of their replicates. When the normal
function of the host is down regulated, phage replication is
correspondingly down shifted. Another reason is the reduced efficiency
during phage infection, either in adsorption rates or viable infection,
which leads to a diminishing of the infection cycle\cite{MIDD00}, i.e.
\begin{equation}
e>f.  \label{ef}
\end{equation}
It might seem as if the model is discrete in the representation of
metabolism since we put in good and poor metabolism by hand. However,
note that the actual metabolism of the community may be somewhere
between good and poor, i.e. a linearly combination, depending on the
probability or fraction to enter either event. Finally, the death of
the phage may be caused by protein cleavage.

The time evolution of the whole community is accessed by random
sampling. In each time step, we have a probability $\mu $ to draw units
in the host layer and a probability $\nu $ to draw units in the phage
layer. In the host layer, we may draw either one unit with probability
$\omega $ or two units with probability $1-\omega $. In the phage
layer, only one unit is drawn. If a combination not listed in Table
\ref{tab:lysis_events} is drawn, such as $EEB$, nothing happens. Thus
all we need to consider are the above events. Using simple
combinatorics, it is straightforward to obtain the probability for the
combinations as follows (Table \ref{tab:lysis_prob}):

\begin{table}[htbp]
\caption{Probabilities for the combinations in the lysis-only model.}
\label{tab:lysis_prob}\centering
\begin{tabular}{cl}
\hline\hline
combination & probability \\[0.5ex] \hline
&  \\[-2.3ex]
$A$ & $\mu \left( 1-\nu \right) \omega \dfrac{m}{K}$ \\[1ex]
$AA$ & $\mu \left( 1-\nu \right) \left( 1-\omega \right) \dfrac{m\left(
m-1\right) }{K\left( K-1\right) }$ \\[1ex]
$AE$ & $\mu \left( 1-\nu \right) \left( 1-\omega \right) \dfrac{2m\left(
K-m\right) }{K\left( K-1\right) }$ \\[1ex]
$AEB$ & $\mu \nu \left( 1-\omega \right) \dfrac{2m\left( K-m\right) }{%
K\left( K-1\right) }\dfrac{n}{W}$ \\[1ex]
$AAB$ & $\mu \nu \left( 1-\omega \right) \dfrac{m\left( m-1\right) }{K\left(
K-1\right) }\dfrac{n}{W}$ \\[1ex]
$B$ & $\left( 1-\mu \right) \nu \dfrac{n}{W}$ \\[2ex] \hline
\end{tabular}%
\end{table}

\noindent
where the factor $2$ accounts the equality in probability for events $AE$ and
$EA$, or $AEB$ and $EAB$.

Thus we obtain the transition matrices for each kind of variation in
the population during each time step, such as $\left\langle T\left(
m+1,n|m,n\right) \right\rangle$, and further the evolution for the
probability in the population with $m$ hosts and $n$ phages at time $t$ $%
P\left( m,n,t\right)$. The reader is referred to Appendix A for
calculational details.

The average of the population is given by summation

\begin{subequations}
\begin{align}
\left\langle m\right\rangle &= \sum_{mn}mP\left( m,n,t\right) ; \\
\left\langle n\right\rangle &= \sum_{mn}nP\left( m,n,t\right) .
\end{align}
\label{mn}
\end{subequations}%

Thus, the time evolution for the population size is

\begin{subequations}
\begin{align}
\dfrac{d\left\langle m\right\rangle }{dt}&= \left\langle T\left(
m+1,n|m,n\right) \right\rangle -\left\langle T\left( m-1,n|m,n\right)
\right\rangle  \notag \\
&\qquad -\left\langle T\left( m-1,n+\alpha -1|m,n\right) \right\rangle  \notag \\
&\qquad -\left\langle T\left( m-1,n+\beta -1|m,n\right) \right\rangle; \\
\dfrac{d\left\langle n\right\rangle }{dt}&=\left( \alpha -1\right)
\left\langle T\left( m-1,n+\alpha -1|m,n\right) \right\rangle  \notag \\
&\qquad + \left( \beta -1\right) \left\langle T\left( m-1,n+\beta -1|m,n\right)
\right\rangle  \notag \\
&\qquad - \left\langle T\left( m,n-1|m,n\right) \right\rangle.
\end{align}
\end{subequations}%

Here we have taken the mean field theory limit and neglected all the
correlations and fluctuations.

Omitting angle-brackets for
simplicity, the equations for the evolution in population are

\begin{subequations}
\begin{align}
\dfrac{dm}{dt} &=rm\left( 1-\dfrac{m}{K}\right) -d_{m}m  \notag \\
&\qquad -\phi mn\left( 1-a_{m}\dfrac{m}{K}\right) ; \\
\dfrac{dn}{dt} &= \gamma \phi mn\left( 1-a_{n}\dfrac{m}{K}\right) -d_{n}n;
\end{align}
\label{lysis1}
\end{subequations}%

\noindent
where
\begin{subequations}
\begin{align}
r &= \dfrac{(2b + d)\mu \left( 1-\nu \right) \left( 1-\omega \right)
}{K};\\
\phi &= \dfrac{2e\mu \nu \left( 1-\omega \right) }{KW};\\
\gamma &= \alpha -1;\\
d_{m} &= \dfrac{(c\omega + d(1 - \omega))\mu \left( 1-\nu \right)}{K};\\
d_{n} &= \dfrac{\left( 1-\mu \right) \nu }{W};\\
a_{m} &= 1 - \dfrac{f}{2e}; \label{am}\\
a_{n} &= 1- \dfrac{\beta f}{2\alpha e}. \label{an}
\end{align}
\label{parameter1}
\end{subequations}

Considering Eq. (\ref{ab}), we notice that Eq. (\ref{am}) (\ref{an})
yield the following relation

\begin{equation}
0<a_{m}<a_{n}<1.
\end{equation}

\noindent
Generally speaking, $a_{m}\neq a_{n}$ unless

\begin{equation}
\alpha =\beta ,
\end{equation}
\noindent which implies that the reproduction numbers under good and
poor metabolism are the same as in Weitz and Dushoff's model.  This
concludes the derivation of the equations for population dynamics from
the individual or microscopic level.

\subsection{Results}

In this section we explore the predictions of the lysis-only model
given by Eq. (\ref{lysis1}).

Let
\begin{subequations}
\begin{align}
t^{\prime } &=\dfrac{rt}{a_{m}}; \\
\phi ^{\prime } &=\dfrac{\phi \gamma K}{r}; \\
d_{n}^{\prime } &=\dfrac{a_{m}d_{n}}{r}; \\
d_{m}^{\prime } &=\dfrac{a_{m}d_{m}}{r}+1-a_{m}; \\
m^{\prime } &=a_{m}\dfrac{m}{K}; \\
n^{\prime } &=\dfrac{a_{m}n}{\gamma K}; \\
a_{n}^{\prime } &=\dfrac{a_{n}}{a_{m}};
\end{align}
\end{subequations}%

We can non-dimensionalize the evolution equations (\ref{lysis1}). Omitting  the primes we obtain
\begin{subequations}
\begin{align}
\dfrac{dm}{dt} &=m\left( 1-m\right) -\phi mn\left( 1-m\right) -d_{m}m; \\
\dfrac{dn}{dt} &=\phi mn\left( 1-a_{n}m\right) -d_{n}n.
\end{align}
\end{subequations}%

\noindent
Setting
\begin{subequations}
\begin{align}
\dfrac{dm}{dt} &=0 \\
\dfrac{dn}{dt} &=0,
\end{align}
\end{subequations}%
we obtain three fixed points. The first is a trivial fixed
point
\begin{subequations}
\begin{align}
m &=0 \\
n &=0,
\end{align}
\end{subequations}%
which is stable when
\begin{equation}
d_{m}>1.
\end{equation}
The second corresponds to the phage extinction phase
\begin{subequations}
\begin{align}
m &=1-d_{m} \\
n &=0,
\end{align}
\end{subequations}%
which is stable when
\begin{equation}
0<d_{m}<1-\dfrac{1}{a_{n}}
\end{equation}%
or
\begin{align}
1-\dfrac{1}{a_{n}} &<d_{m}<1 \\
\dfrac{\phi }{d_{n}} &<\dfrac{1}{\left( 1-d_{m}\right) \left[ 1-a_{n}\left(
1-d_{m}\right) \right] }.
\end{align}

The last is the coexistence of hosts and phages
\begin{subequations}
\begin{align}
m &=\rho \\
n &=\dfrac{1}{\phi }\left( 1+\dfrac{d_{m}}{\rho -1}\right) ,
\end{align}
\end{subequations}%
where $\rho $ is a root of
\begin{equation}
a_{n}\phi \rho ^{2}-\phi \rho +d_{n}=0.
\end{equation}
The coexistence phase comes into existence and will be stable when
\begin{align}
\dfrac{\phi }{d_{n}} &\geq 4a_{n} \\
d_{m} &<1-\rho .
\end{align}
The stability of the fixed points are governed by the Jacobian
\begin{equation}
\begin{pmatrix}
\left( 1-2m\right) \left( 1-\phi n\right) -d_{m} & -\phi m\left( 1-m\right)
\\
\phi n\left( 1-2a_{n}m\right) & \phi m\left( 1-a_{n}m\right) -d_{n}%
\end{pmatrix}%
\end{equation}
to the equations
\begin{subequations}
\begin{align}
m\left( 1-m\right) -\phi mn\left( 1-m\right) -d_{m}m &=0  \label{fp1} \\
\phi mn\left( 1-a_{n}m\right) -d_{n}n &=0.
\end{align}
\end{subequations}%

\begin{figure}[htbp]
\includegraphics[width=0.99\columnwidth]{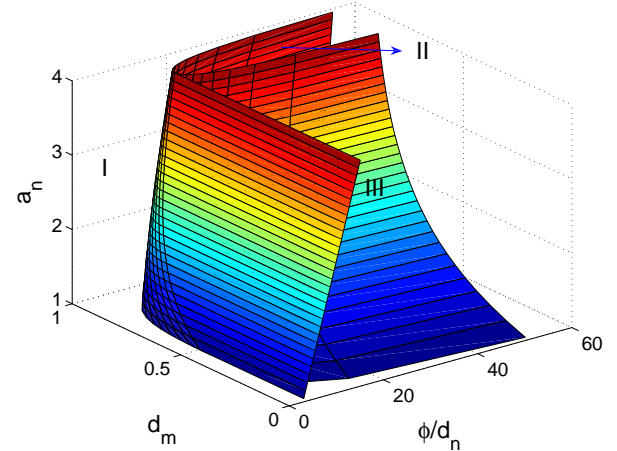}\newline
\caption{(Color online) Three dimensional phase diagram for the lysis-only
model. Region I depends on the initial conditions to flow to the phage
extinction or coexistence fixed point. Region II and III are basins of
attraction for coexistence and phage extinction fixed points, respectively.}
\label{pd1}
\end{figure}

Thus we obtain the three-dimensional phase diagram plotted in Fig. \ref{pd1}. The
basin of attraction for the trivial case is not plotted. Region II is the
basin of attraction for coexistence fixed point only while region III is
that for the phage extinction. Region I will either go to coexistence or
phage extinction, depending on the initial conditions.

\subsection{Discussion}

\begin{figure}[htbp]
\includegraphics[width=0.99\columnwidth]{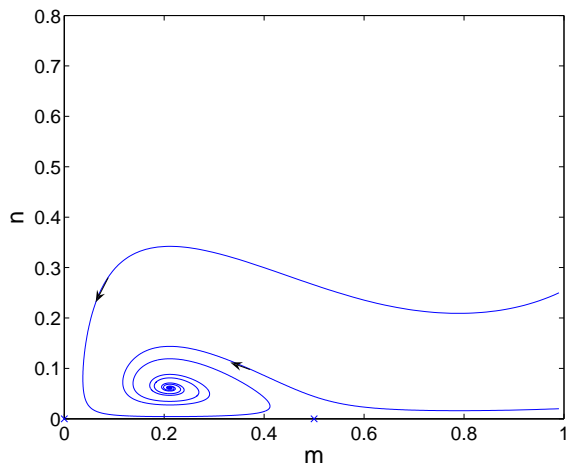}\newline
\caption{(Color online) Flow diagram for $a_{n}=1,\protect\phi
=5,d_{n}=1,d_{m}=0.1.$ ``$\times $" denotes saddle points and ``$\cdot $" is
for stable fixed points.}
\label{joshua}
\end{figure}

\begin{figure}[htbp]
\includegraphics[width=0.99\columnwidth]{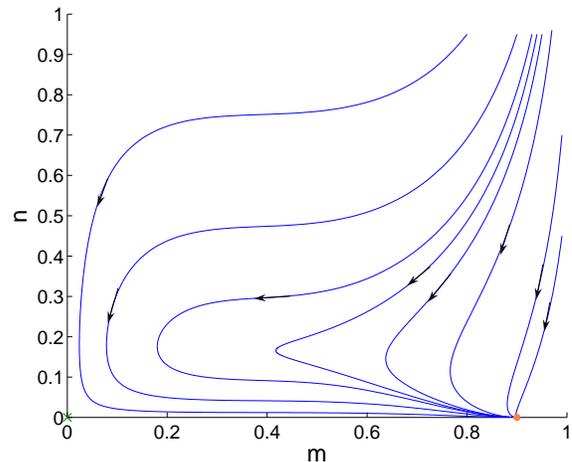}\newline
\caption{(Color online) Flow diagram for $a_{n}=1.3,\protect\phi %
=5,d_{n}=1,d_{m}=0.1.$ ``$\times $" denotes saddle points and ``$\cdot $" is
for stable fixed points.}
\label{joshua2}
\end{figure}

As we can see, the bottom plane in Fig. \ref{pd1} corresponds to the phase
diagram in\ Weitz and Dushoff's model, where $a_{n}=1.$ When
\begin{equation}
\alpha >\beta
\end{equation}
leading to
\begin{equation}
a_{n}>1,
\end{equation}
there is a shift in the phase diagram with a rapid shrinkage of the
basin of attraction for region II, where any initial condition flows to
the coexistence phase. The boundary between region I and III also moves
to larger $\dfrac{\phi }{d_{n}},$ which implies that the more the good
and poor metabolisms differ from each other in the progeny number, the
easier the phages are driven out of the system. In order to see the
effect of the phase shift more clearly, let us tune $a_{n}=1.3$ while
keeping all the other parameters as those in Fig. 2 (I) in Weitz and
Dushoff's paper\cite{WEIT08b} (Fig. \ref{joshua}). When $a_{n}=1,$ there
is a neutral fixed point for coexistence. However, such a fixed point
disappears (Fig. \ref{joshua2}) when $a_{n}=1.3.$ The flow diagrams are
generated by 4$ ^{\text{th}}$ order Runge-Kutta method.

In summary, we have obtained Weitz and Dushoff's model by detailed derivation
from the individual or microscopic level and found a small shift in the phase diagram.
Such a shift, as we see, can be observed experimentally by the onset of
coexistence for the two species.

\section{Lysogeny-lysis model}
\label{lysogeny-lysis}

\subsection{Derivation of the population dynamics from an individual-level model}

Now we extend the lysis-only model above to incorporate lysogeny and
investigate the important role of lysogeny in host-phage dynamics. Now
there are three types of organism in the community. They are ``healthy"
hosts, which have no integration of phage genes, lysogens, and free
phages, which live outside bacteria membranes. We will label ``healthy"
hosts, lysogens and free phages by $A$, $D$, and $B$, respectively,
with population sizes $m$, $s$ and $n$.  For the same reasons as in the
lysis-only model, hosts and phages are thought of as being confined in
different layers characterized by different carrying capacities. Hence
both ``healthy" hosts and lysogens are in the host layer with a total
carrying capacity $K$. The empty sites in the host layer are
denoted by $E$ and their number is $K-m-s$. In the phage layer, the empty
sites are labelled by $\phi$ as before.

The incorporation of lysogens brings us more microscopic events. There are
two pathways after phage infections: lysis and lysogeny. Immediate lysis
for temperate phages is the same
as virulent ones, which has been characterized by events in the previous
section. Lysogeny is an option only for temperate phages, which will be
investigated in detail here. First, there should be an event corresponding to
lysogen formation, i.e. a phage integrates its DNA into the genome of the
host and turns itself into a prophage. Second, lysogens will survive, replicate
and die as ``healthy" hosts. Last, environments might trigger prophage
induction, which lyses the lysogen and releases the prophages inside.
In all, there are eighteen microscopic events, which are listed
in Table \ref{tab:lysogeny_events}.

\begin{table}[htbp]
\caption{Microscopic events in the lysogeny-lysis model.}
\label{tab:lysogeny_events}\centering
\begin{tabular}{lc}
\hline\hline
description & symbols \\[0.5ex] \hline
&  \\[-2.6ex]
birth of host & $AE\overset{b}{\rightarrow }AA$ \\[0.5ex]
& $DE\overset{b}{\rightarrow }DD$ \\[0.5ex]
death of host due to longevity & $A\overset{c}{\rightarrow }E$ \\[0.5ex]
& $D\overset{c}{\rightarrow }E$ \\[0.5ex]
death of host due to crowding & $AA\overset{d}{\rightarrow }AE$ \\[0.5ex]
& $DD\overset{d}{\rightarrow }DE$ \\[0.5ex]
& $AD\overset{\frac{1}{2}d}{\rightarrow }DE$ \\[0.5ex]
& $AD\overset{\frac{1}{2}d}{\rightarrow }AE$ \\[0.5ex]
host-phage interactions &  \\[0.5ex]
$\cdot $ lysis under good metabolism & $AEB\overset{e}{\rightarrow }EE\alpha
B$ \\[0.5ex]
$\cdot $ lysis under poor metabolism & $AAB\overset{f}{\rightarrow }EA\beta
B $ \\
[0.5ex] & $ADB\overset{f}{\rightarrow }ED\beta B$ \\[0.5ex]
$\cdot $ lysogeny under good metabolism & $AEB\overset{h}{\rightarrow }DE$ \\%
[0.5ex]
$\cdot $ lysogeny under poor metabolism & $AAB\overset{k}{\rightarrow }DA$ \\%
[0.5ex]
& $ADB\overset{k}{\rightarrow }DD$ \\[0.5ex]
prophage induction &  \\[0.5ex]
$\cdot $ under good metabolism & $DE\overset{p}{\rightarrow }EE\alpha B$ \\%
[0.5ex]
$\cdot $ under poor metabolism & $DD\overset{q}{\rightarrow }DE\beta B$ \\%
[0.5ex]
& $DA\overset{q}{\rightarrow }AE\beta B$ \\[0.5ex]
death of free phage & $B\overset{g}{\rightarrow }\phi $ \\[1ex] \hline
\end{tabular}%
\end{table}

Here $b$, $c$, $d$, $e$, $f$, $g$, $h$, $k$, $p$ and $q$ are constant
reaction rates. $\alpha $ and $\beta $ are phage reproduction numbers
under good and poor metabolisms, respectively. Although prophage
induction enhances the survival ability for lysogens in several
ways, such as suppressing the latter's metabolism\cite {PAUL08} through
down-regulation\cite{CHEN05}, for simplicity we have assumed the same birth and death
rates for ``healthy" hosts and lysogens. We have the condition
\begin{equation}
\alpha >\beta,
\end{equation}
as before. Furthermore, there are the following advantages under better
metabolism: more successful and effective infection (Eq. (\ref{eh1})), greater
possibility to lyse the host (Eq. (\ref{eh2})), and faster prophage release
(Eq. (\ref{pq})). Since lysis is controlled by lytic repressor CI dimers
while lysogeny is regulated by CII, we do not expect any special relationship between
$e$ and $f$, and $p$ and $q$.  These advantages can be expressed
mathematically by the following inequalities:

\begin{subequations}
\begin{align}
e+h &>f+k,  \label{eh1}\\
\dfrac{e}{h} &>\dfrac{f}{k},  \label{eh2} \\
p &>q.  \label{pq}
\end{align}
\end{subequations}
We draw events from the two layers the same way as in the lysis-only
model and this results in the probabilities shown in Table \ref{tab:lysogeny_prob}.

From these events, we obtain the following evolution equations for all the three
species after the calculations provided in Appendix B:

\begin{widetext}
\begin{subequations}
\allowdisplaybreaks
\begin{align}
\dfrac{dm}{dt}&=rm\left( 1-\dfrac{m+s}{K}\right) -d_{1}m-\phi
_{1}mn\left\{ 1-\dfrac{1}{K}\left[ \left( 1-a_{1}\right) m+\left(
1-2a_{1}\right) s\right]
\right\} ; \label{lysogeny_dm1}\\
\dfrac{ds}{dt}&=rs\left( 1-\dfrac{m+s}{K}\right) -d_{1}s+\phi
_{2}mn\left\{
1-\dfrac{1}{K}\left[ \left( 1-a_{21}\right) m+\left( 1-2a_{21}\right) s%
\right] \right\}  \nonumber\\
&\qquad -d_{2}s\left\{ 1-\dfrac{1}{K}[\left( 1-2a_{22}\right) m+\left(
1-a_{22}\right) s]\right\} ; \label{lysogeny_ds1}\\
\dfrac{dn}{dt}&=\left[ \left( \alpha -1\right) \phi _{1}-\alpha \phi _{2}%
\right] mn\left\{ 1-\dfrac{1}{K}\left[ \left( 1-a_{31}\right)
m+\left(
1-2a_{31}\right) s\right] \right\}  \nonumber\\
&\qquad +\alpha d_{2}s\left\{ 1-\dfrac{1}{K}\left[ \left( 1-2a_{32}\right)
m+\left( 1-a_{32}\right) s\right] \right\}
-d_{3}n;\label{lysogeny_dn1}
\end{align}
\label{lysogeny1}
\end{subequations}
\end{widetext}%

\noindent
where
\begin{subequations}
\allowdisplaybreaks
\begin{align}
r &= \dfrac{(2b + d) \mu \left( 1-\nu \right) \left( 1-\omega \right)
}{K}; \\
d_{1} &= \dfrac{(c\omega + d (1 - \omega))\mu \left( 1-\nu \right)}{K}; \\
d_{2} &= \dfrac{2p\mu \left( 1-\nu \right) \left( 1-\omega \right)
}{K}; \\
d_{3} &= \dfrac{\left( 1-\mu \right) \nu }{W}; \\
\phi_{1} &= \dfrac{2(e + h)\mu \nu \left( 1-\omega \right) }{KW}; \\
\phi_{2} &= \dfrac{2h\mu \nu \left( 1-\omega \right) }{KW}; \\
a_{1} &= \dfrac{f + k}{2(e + h)}; \\
a_{21} &= \dfrac{k}{2h}; \\
a_{22} &= \dfrac{q}{2p}; \\
a_{31} &= \dfrac{\beta f-k}{2(\alpha e - h)}; \\
a_{32} &= \dfrac{\beta q}{2\alpha p}.
\end{align}
\label{parameter2}
\end{subequations}

\begin{table}[htbp]
\caption{Probabilities for the combinations in the lysogeny-lysis model.}
\label{tab:lysogeny_prob}\centering
\begin{tabular}{cl}
\hline\hline
combination & probability \\[0.5ex] \hline
&  \\[-2.3ex]
$AE$ & $\mu \left( 1-\nu \right) \left( 1-\omega \right) \dfrac{2m\left(
K-m-s\right) }{K\left( K-1\right) }$ \\[1.3ex]
$DE$ & $\mu \left( 1-\nu \right) \left( 1-\omega \right) \dfrac{2s\left(
K-m-s\right) }{K\left( K-1\right) }$ \\[1.3ex]
$A$ & $\mu \left( 1-\nu \right) \omega \dfrac{m}{K}$ \\[1.3ex]
$D$ & $\mu \left( 1-\nu \right) \omega \dfrac{s}{K}$ \\[1.3ex]
$AA$ & $\mu \left( 1-\nu \right) \left( 1-\omega \right) \dfrac{m\left(
m-1\right) }{K\left( K-1\right) }$ \\[1.3ex]
$DD$ & $\mu \left( 1-\nu \right) \left( 1-\omega \right) \dfrac{s\left(
s-1\right) }{K\left( K-1\right) }$ \\[1.3ex]
$AD$ & $\mu \left( 1-\nu \right) \left( 1-\omega \right) \dfrac{2ms}{K\left(
K-1\right) }$ \\[1.3ex]
$AEB$ & $\mu \nu \left( 1-\omega \right) \dfrac{2m\left( K-m-s\right) }{%
K\left( K-1\right) }\dfrac{n}{W}$ \\[1.3ex]
$AAB$ & $\mu \nu \left( 1-\omega \right) \dfrac{m\left( m-1\right) }{K\left(
K-1\right) }\dfrac{n}{W}$ \\[1.3ex]
$ADB$ & $\mu \nu \left( 1-\omega \right) \dfrac{2ms}{K\left( K-1\right) }%
\dfrac{n}{W}$ \\[1.3ex]
$B$ & $\left( 1-\mu \right) \nu \dfrac{n}{W}$ \\[2ex] \hline
\end{tabular}%
\end{table}

We note that

\begin{align}
\phi _{2} &<\phi _{1}; \\
0 &<a_{1},a_{21},a_{22},a_{31},a_{32}<1; \\
a_{32} &<a_{22}.
\end{align}

We also notice some kind of symmetry in the correction terms such as ``$%
1-a_{1} $" and ``$1-2a_{1}$". $a_{1}$ originates from the poor metabolism of
hosts $A$, which indirectly downshifts the efficiency of phage infection and
synthesis. In equation (\ref{lysogeny_dm1}), ``$a_{1}$" comes from the event $%
AAB\overset{f}{\rightarrow }EA\beta B,$ while ``$2a_{1}$" is from $ADB\overset%
{f}{\rightarrow }ED\beta B.$ The factor ``$2$" appears since ``$AD$" is the
same as ``$DA$".

Considering

\begin{equation}
\alpha \gg 1,
\end{equation}

\noindent
for example,

\begin{equation}
\alpha \approx 100
\end{equation}

\noindent
for lambda phage\cite{PTAS04}, we approximate

\begin{equation}
\left( \alpha -1\right) \phi _{1}-\alpha \phi _{2}\approx \alpha \left( \phi
_{1}-\phi _{2}\right) .
\end{equation}

Hence equation (\ref{lysogeny_dn1}) can be simplified as

\begin{widetext}
\begin{equation}
\dfrac{dn}{dt}=\alpha \left( \phi _{1}-\phi _{2}\right) mn\left\{ 1-\dfrac{1%
}{K}\left[ \left( 1-a_{31}\right) m+\left( 1-2a_{31}\right) s\right]
\right\} +\alpha d_{2}s\left\{ 1-\dfrac{1}{K}\left[ \left( 1-2a_{32}\right)
m+\left( 1-a_{32}\right) s\right] \right\} -d_{3}n.
\label{lysogeny-simplification}
\end{equation}
\end{widetext}

\subsection{Results}

In this section, we explore the predictions of the lysogeny-lysis model
given by equations (\ref{lysogeny_dm1}), (\ref{lysogeny_ds1}) and (\ref{lysogeny-simplification}).

Let
\begin{subequations}
\allowdisplaybreaks
\begin{align}
t^{\prime } &=rt; \\
\phi _{1}^{\prime } &=\dfrac{\alpha \phi _{1}K}{r}; \\
\phi _{2}^{\prime } &=\dfrac{\alpha \phi _{2}K}{r}; \\
d_{1}^{\prime } &=\dfrac{d_{1}}{r}; \\
d_{2}^{\prime } &=\dfrac{d_{2}}{r}; \\
d_{3}^{\prime } &=\dfrac{d_{3}}{r}; \\
m^{\prime } &=\dfrac{m}{K}; \label{mprime}\\
s^{\prime } &=\dfrac{s}{K}; \label{sprime}\\
n^{\prime } &=\dfrac{n}{\alpha K}; \label{nprime}
\end{align}
\end{subequations}%
and omitting the primes, the equations after non-dimensionalization
become
\begin{widetext}
\begin{subequations}
\allowdisplaybreaks
\begin{align}
\dfrac{dm}{dt}&=m\left( 1-m-s\right) -d_{1}m-\phi _{1}mn\left[
1-\left(
1-a_{1}\right) m-\left( 1-2a_{1}\right) s\right] ; \\
\dfrac{ds}{dt}&=s\left( 1-m-s\right) -d_{1}s+\phi _{2}mn\left[
1-\left( 1-a_{21}\right) m-\left( 1-2a_{21}\right) s\right]
-d_{2}s\left[ 1-\left(
1-2a_{22}\right) m-\left( 1-a_{22}\right) s\right] ; \\
\dfrac{dn}{dt}&=\left( \phi _{1}-\phi _{2}\right) mn\left[ 1-\left(
1-a_{31}\right) m-\left( 1-2a_{31}\right) s\right] +d_{2}s\left[
1-\left( 1-2a_{32}\right) m-\left( 1-a_{32}\right) s\right] -d_{3}n.
\end{align}
\label{lysogeny_fp2}
\end{subequations}
\end{widetext}%

Formally, the fixed points can be solved by requiring that
\begin{subequations}
\allowdisplaybreaks
\begin{align}
\dfrac{dm}{dt} &=0; \\
\dfrac{ds}{dt} &=0; \\
\dfrac{dn}{dt} &=0.
\end{align}
\end{subequations}%
However, we can only obtain four fixed points analytically. The
first is the trivial case for the extinction of all the species
\begin{subequations}
\begin{align}
m &=0; \\
s &=0; \\
n &=0.
\end{align}
\end{subequations}%
The second is the ``healthy" host extinction fixed point
\begin{subequations}
\begin{align}
m &=0; \\
s &=\dfrac{1-d_{1}-d_{2}}{1-d_{2}\left( 1-a_{22}\right) }; \\
n &=\dfrac{d_{2}}{d_{3}}s\left[ 1-\left( 1-a_{32}\right) s\right] .
\end{align}
\end{subequations}%
The third is the ``healthy" host only fixed point
\begin{subequations}
\allowdisplaybreaks
\begin{align}
m &=1-d_{1}; \\
s &=0; \\
n &=0.
\end{align}
\end{subequations}%
The last is the lysogen extinction
\begin{subequations}
\allowdisplaybreaks
\begin{align}
m &=\dfrac{1}{1-a_{21}}; \\
s &=0; \\
n &=\dfrac{1-m-d_{1}}{\phi _{1}\left[ 1-\left( 1-a_{1}\right) m\right] };
\end{align}
\end{subequations}%
whose existence requires that
\begin{equation}
\left( \phi _{1}-\phi _{2}\right) \left( a_{31}-a_{21}\right) =d_{3}\left(
1-a_{21}\right) ^{2}.
\end{equation}

\begin{figure}[tbp]
\includegraphics[width=0.99\columnwidth]{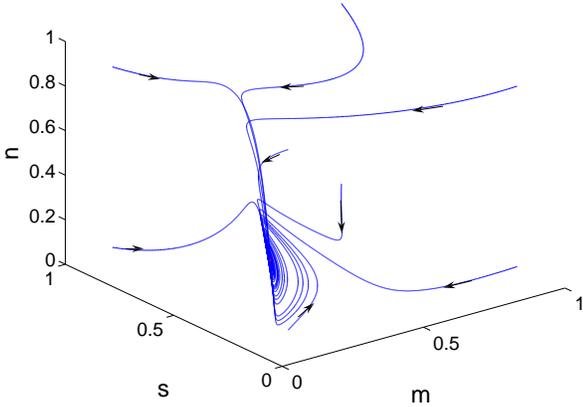}\newline
\caption{(Color online) In the lysogeny-lysis model, a stable fixed point
for the coexistence of all the three species. The parameters are $\protect%
\phi _{1}=1,\protect\phi %
_{2}=0.8,d_{1}=0.5,d_{2}=0.49,d_{3}=0.1,a_{1}=a_{21}=a_{31}=0.1,a_{22}=a_{32}=0.5.
$}
\label{lysogeny_fp1}
\end{figure}

The more interesting coexistence of all the three species is hard to solve
analytically since the order of the equations
\begin{widetext}
\begin{subequations}
\begin{align}
m\left( 1-m-s\right) -d_{1}m-\phi _{1}mn\left[ 1-\left(
1-a_{1}\right)
m-\left( 1-2a_{1}\right) s\right] &=0; \\
s\left( 1-m-s\right) -d_{1}s+\phi _{2}mn\left[ 1-\left(
1-a_{21}\right) m-\left( 1-2a_{21}\right) s\right] -d_{2}s\left[
1-\left( 1-2a_{22}\right)
m-\left( 1-a_{22}\right) s\right] &=0; \\
\left( \phi _{1}-\phi _{2}\right) mn\left[ 1-\left( 1-a_{31}\right)
m-\left( 1-2a_{31}\right) s\right] +d_{2}s\left[ 1-\left(
1-2a_{32}\right) m-\left( 1-a_{32}\right) s\right] -d_{3}n &=0;
\end{align}
\end{subequations}
\end{widetext}%
is too high. Using a 4$^{\text{th}}$ order Runge-Kutta method, we found numerically a
stable fixed point, shown in Fig. \ref{lysogeny_fp1}.

\subsection{Discussion}

\begin{figure}[tbp]
\includegraphics[width=0.99\columnwidth]{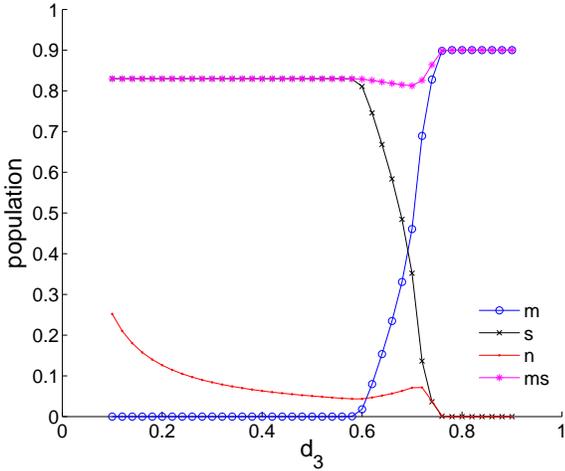}\newline
\caption{(Color online) The population of the community with increasing
phage mortality rate $d_{3}.$ ``$ms$" indicates the sum of $m$ and $s$.}
\label{lysogeny_ph1}
\end{figure}

As shown in Eq. (\ref{lysogeny_fp2}), there are, in total, ten
parameters so that the phase space is difficult to visualize. We have
studied the general trend of the transition between phases, starting
with the dependence on phage mortality rate $d_3$. In Fig.
\ref{lysogeny_ph1}, it is shown that when the phage mortality rate is
low, the systems flows into a ``healthy" host extinction phase. The
phage population decreases with increase in the phage mortality rate,
which is very reasonable physically. For intermediate values of $d_3$,
there is coexistence for all the three species, while for large values
of $d_3$, the only survival is ``healthy" host, where all phages die
out quickly out, leading to the extinction of lysogens.

\begin{figure}[tbp]
\includegraphics[width=0.99\columnwidth]{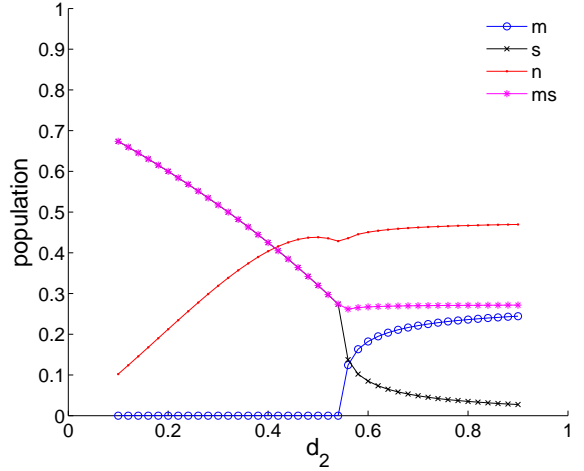}\newline
\caption{(Color online) The population for the community with the increase
in the lysis rate $d_{2}.$}
\label{lysogeny_ph2}
\end{figure}

We show the trend of the population with increasing lysis rate
$d_{2}$ in Fig. \ref{lysogeny_ph2}. The phage prospers with the
increase in the lysis rate, while the lysogen diminishes. The peak in
the phage population appears when there is a balance in the number of
lysogens available to lyse and the lysis rate. When the lysis rate is
beyond the threshold at 0.54, lysogen number falls dramatically and
there is a proliferation of ``healthy" hosts. The total host population
is roughly the same afterwards while the phage population upshifts a
little with the increase in the ``healthy" host available to infect but
does not change further when the ratio between ``healthy" hosts and
lysogens converges.

We have studied the effect of host mortality rate in Fig. \ref%
{lysogeny_ph3}. Obviously the total host population will fall
monotonically when the hosts are more likely to die. We draw attention
to the interesting peak in the phage population. When the host
mortality rate is low, the phage population is suppressed due to the
overcrowding of the lysogens, which degrades the metabolism and hence
the infection and synthesis of phages. When the host mortality rate is
high, on the other hand, the phages have insufficient hosts to
infect and their population also declines.

\begin{figure}[tbp]
\includegraphics[width=0.99\columnwidth]{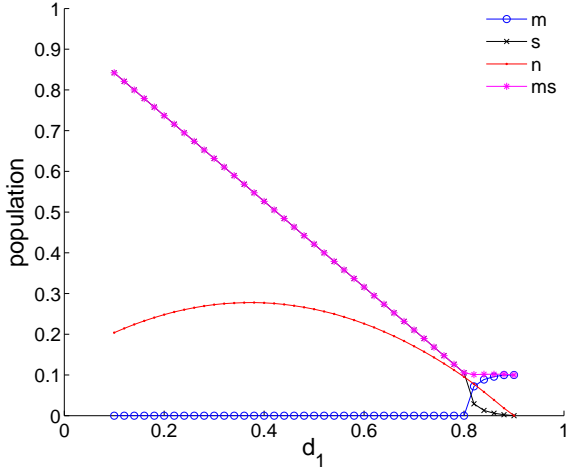}\newline
\caption{(Color online) The population for the community with the increase
in the host mortality rate $d_{1}.$}
\label{lysogeny_ph3}
\end{figure}

\subsection{Existence of a limit cycle}

\begin{figure}[htbp]
\includegraphics[width=0.99\columnwidth]{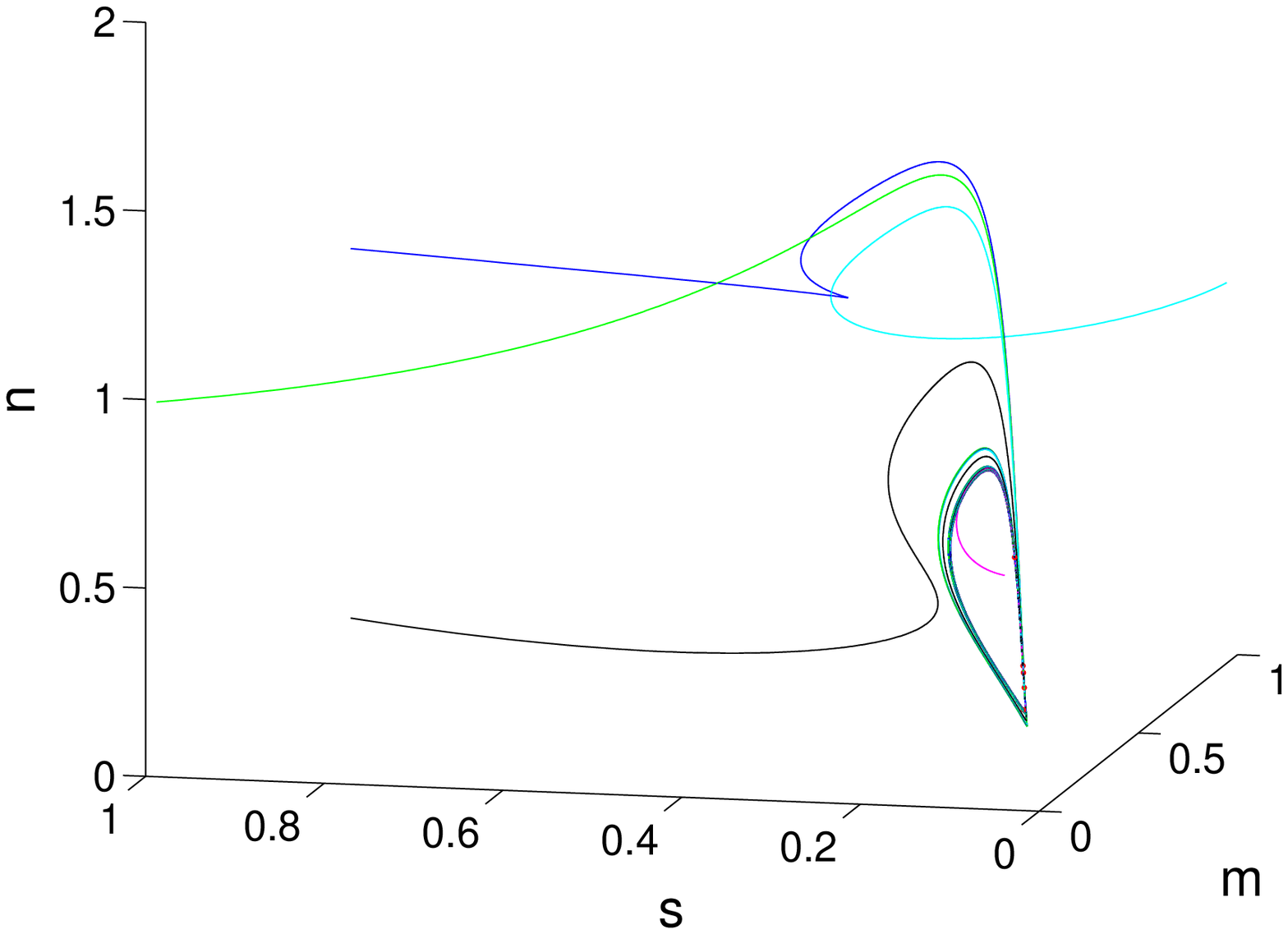}\newline
\caption{(Color online) A limit cycle in the flow diagram for different
initial conditions with parameters $\protect\phi _{1}=1,\protect\phi %
_{2}=0.8,d_{1}=0.5,d_{2}=0.49,d_{3}=0.03,a_{1}=a_{21}=a_{31}=0.1,a_{22}=a_{32}=0.5.
$}
\label{limit_cycle1}
\end{figure}

\begin{figure}[htbp]
\includegraphics[width=0.99\columnwidth]{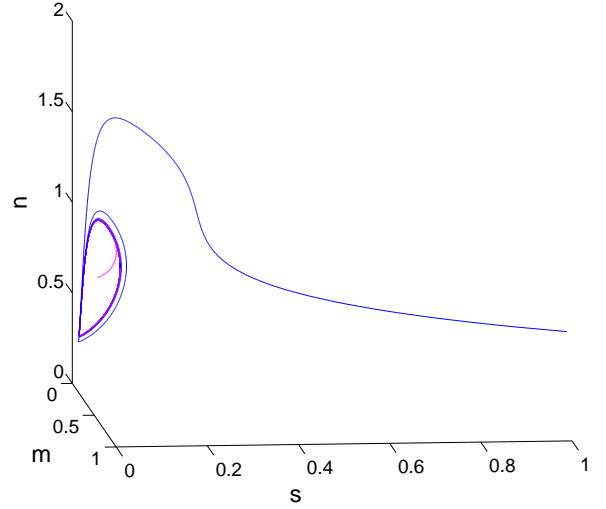}\newline
\caption{(Color online) A limit cycle in the flow diagram with different
initial conditions for parameters $\protect\phi _{1}=1,\protect\phi %
_{2}=0.8,d_{1}=0.5,d_{2}=0.49,d_{3}=0.03,a_{1}=a_{21}=a_{31}=0.1,a_{22}=a_{32}=0.5
$. The limit cycle is in a curved space. The blue curve initiated outside
the cycle flows in while the red one from inside flows out.}
\label{limit_cycle2}
\end{figure}%

\begin{figure}[htbp]
\includegraphics[width=0.99\columnwidth]{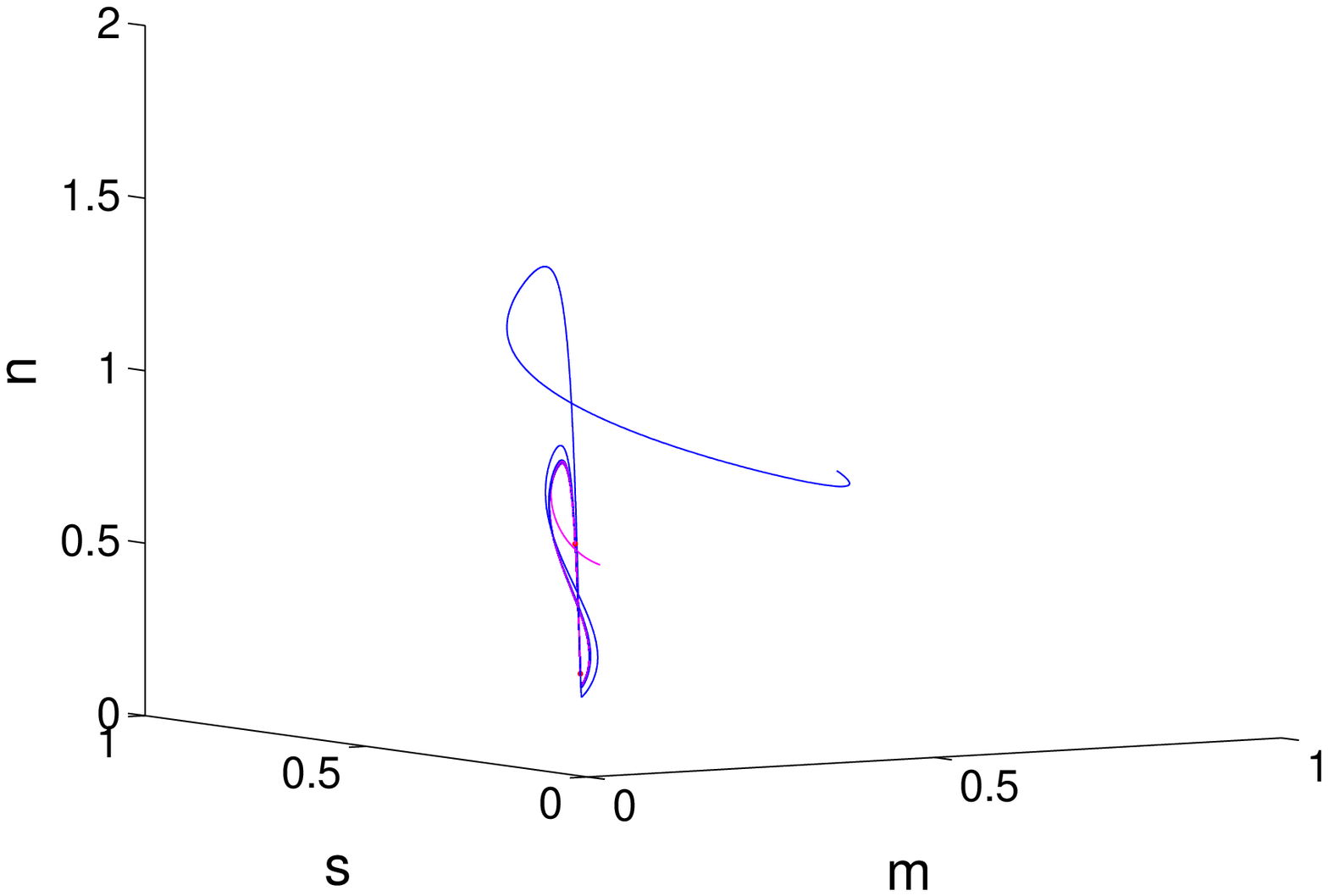}\newline
\caption{(Color online) A limit cycle in the flow diagram with different
initial conditions for parameters $\protect\phi _{1}=1,\protect\phi %
_{2}=0.8,d_{1}=0.5,d_{2}=0.49,d_{3}=0.03,a_{1}=a_{21}=a_{31}=0.1,a_{22}=a_{32}=0.5
$. The limit cycle is in a curved space.}
\label{limit_cycle3}
\end{figure}%

\begin{figure}[htbp]
\includegraphics[width=0.8\columnwidth]{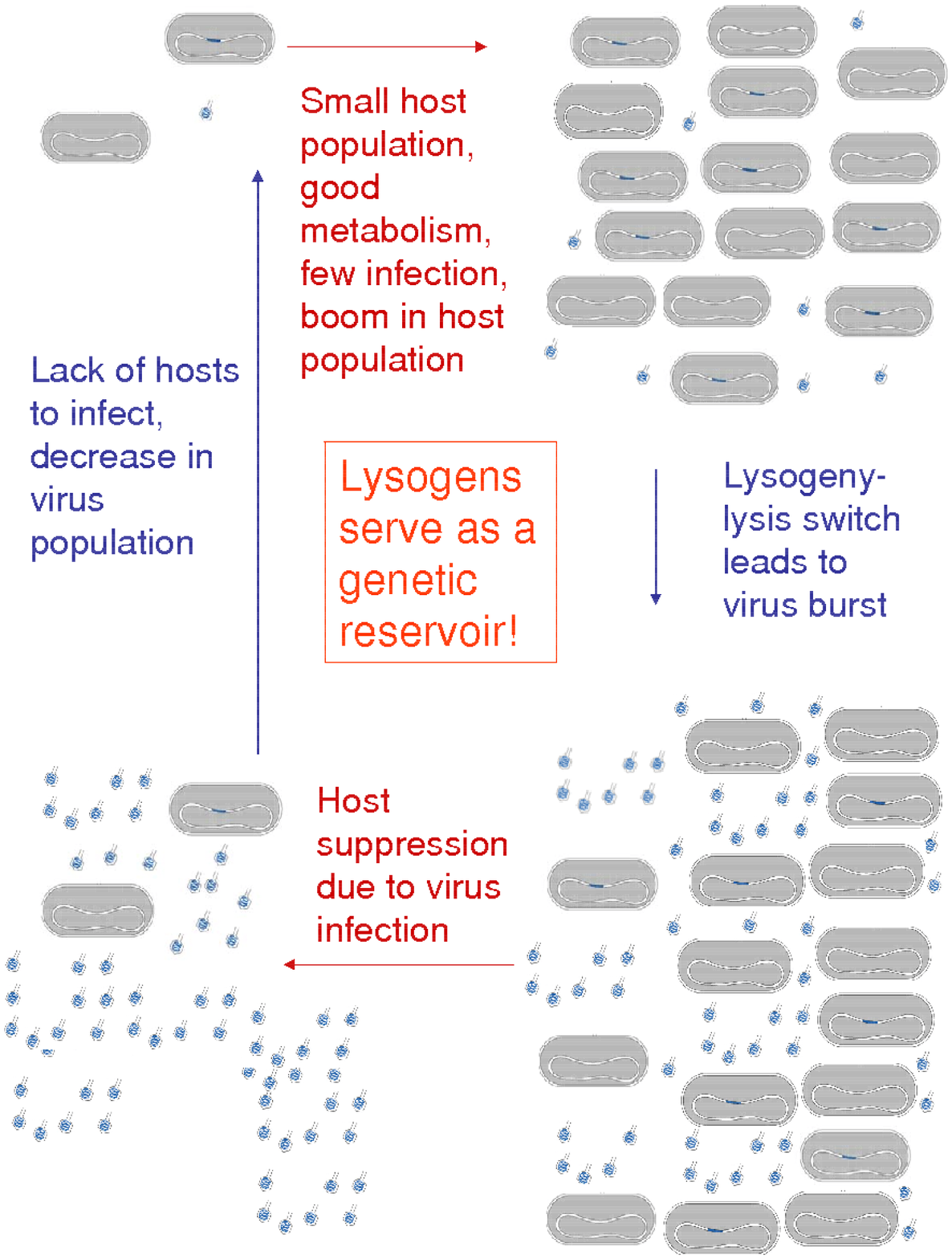}\newline
\caption{(Color online) Cartoon explanation for the limit cycle.
When the population for host and phages are both small, the host will
enjoy a boom because of good metabolism and little phage infection.
Meanwhile prophages replicate with the fast reproduction of lysogens.
Once the lysogeny-lysis switch is triggered, the destruction of lysogens
will yield a huge virus burst. Then ``healthy" host will encounter intensive
phage infection and hence be suppressed. When most of the host die out,
phage population shrinks quickly due to lack of infection. In this way,
a cycle forms.}
\label{cartoon}
\end{figure}%

We have noticed that the dynamics exhibits a limit cycle for some
combination of parameters (Fig. \ref{limit_cycle1}).  In this section,
we describe our numerical evidence for this assertion and present a
plausible physical interpretation of our finding. In order to verify
that it is a limit cycle instead of some unexpected slowing down near a
putative stable or neutral fixed point, we have chosen an initial
condition located inside the conjectured limit cycle. If there is, in
fact, no real limit cycle, the dynamics will flow inwards no matter how
slow it will be. However, as we can see in Fig. \ref{limit_cycle2}, the
flow indicated by the red curve flows out. Hence we have observed in
the flow diagram an oscillation in the population for all the three
species. If we inspect neighboring time steps, it appears that the
convergence is slow, since the deviation from step to step is very
small.
However, on longer time scales, we can see that the convergence is an
illusion. Moreover, tilting the view angle, we see that the limit cycle
is in some curved space instead of a single plane in Fig. \ref
{limit_cycle3}. In order to investigate the emergence of the limit
cycle, we have scanned part of the parameter space. For example,
there is a stable coexistence fixed point for $d_1 > 0.41$
while $\phi_1  = 1, \phi_2 = 0.8, d_2 = 0.9, d_3 = 0.048, a_1 =%
a_{21} = a_{31} = 0.1$, and $ a_{22} = a_{32} = 0.5$. However, the
above fixed point becomes unstable if $d_1 < 0.41$ leading to the limit
cycle. As we see it, such an oscillation for the population in the
community is a manifestation of the role of lysogens (Fig. \ref{cartoon}).
When the population for host and phages are both small, the host will
enjoy a boom because of good metabolism and little phage infection.
Meanwhile prophages replicate with the fast reproduction of lysogens.
Once the lysogeny-lysis switch is triggered, the destruction of lysogens
will yield a huge virus burst. Then ``healthy" host will encounter intensive
phage infection and hence be suppressed. When most of the host die out,
phage population shrinks quickly due to lack of infection. In this way,
a cycle forms.
Integrating its DNA into the genome of a lysogen, a prophage is sheltered
although it is temporarily dormant in the sense of viral
infection. Such a stage assists prophages to survive
demanding environmental conditions and provides an opportunity to
resurrect the population when there are abundant ``healthy" hosts. Thus
lysogens are perfect genetic reservoirs for phages for potential future
burst\cite{GOLD07,PAUL08}.

\section{Stochastic simulation}
\label{stochastic-simulation} Up to now, all the calculations above
were carried out within the scope of mean field theory. As a next step,
it is important to see to what extent such predictions are disturbed by
demographic fluctuations, and especially whether the limit cycle in the
lysogeny-lysis model is stable. A second goal of this section is to
link the parameters the parameters in our model to those which could
characterize real experiments. In this section, we perform stochastic
simulations using the Gillespie's algorithm\cite{GILL76,GILL77}, which is a
very efficient strategy to simulate chemical reactions.  The reaction
rates ($b$, $c$, $d$,$e$, $f$ and $g$ in Table \ref{tab:lysis_events},
and $b$, $c$, $d$, $e$, $f$, $g$, $h$, $k$, $p$ and $q$ in Table
\ref{tab:lysogeny_events}) are interpreted as average probability
rates for the occurrence of the corresponding reactions in line with
the Gillespie algorithm, where the effect of draw probability is
incorporated automatically.

In the lysis-only model, the map between the two sets of parameters
for reactions is

\begin{subequations}
\allowdisplaybreaks
\begin{align}
\widetilde{b} &= bK; \\
\widetilde{c} &= c; \\
\widetilde{d} &= \frac{1}{2}dK; \\
\widetilde{e} &= eK; \\
\widetilde{f} &= \frac{1}{2}fK; \\
\widetilde{g} &= g;
\end{align}
\end{subequations}%

\noindent where tilde is used to indicate the probability rates
in the Gillespie algorithm. Since there are more degrees of freedom in
choosing microscopic event rates, different stochastic simulations
may map into the same mean field phase diagram.

Our main interest is to explore the mean field limit cycle in the
lysogeny-lysis model. We keep employing the tilde symbol
to label the probability rates in the Gillespie sense and the map is

\begin{subequations}
\begin{align}
\widetilde{b} &= bK; \\
\widetilde{c} &= c; \\
\widetilde{d} &= \frac{1}{2}dK; \\
\widetilde{e} &= eK; \\
\widetilde{f} &= \frac{1}{2}fK; \\
\widetilde{h} &= hK; \\
\widetilde{k} &= \frac{1}{2}kK; \\
\widetilde{p} &= pK; \\
\widetilde{q} &= \frac{1}{2}qK; \\
\widetilde{g} &= g.
\end{align}
\end{subequations}%

\begin{figure}[tbp]
\includegraphics[width=1.05\columnwidth]{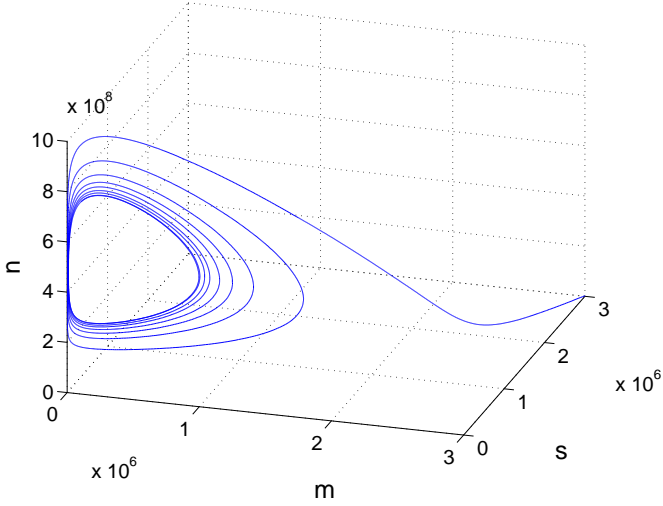}\newline
\caption{A limit cycle in the phase space with parameters in the
Gillespie algorithm $\widetilde{b} = 0.4, \widetilde{c} = 0.1,%
\widetilde{d} = 0.2, \widetilde{e} = 1.2\times10^{-10},%
\widetilde{f} = 1.2\times10^{-11}, \widetilde{g} = 0.018,%
\widetilde{h} = 4.8\times10^{-10}, \widetilde{k} =%
4.8\times10^{-11}, \widetilde{p} = 0.54$, and $\widetilde{q} =%
0.27$.}
\label{limit_cycle_g} \
\end{figure}

\begin{figure}[tbp]
\includegraphics[width=0.99\columnwidth]{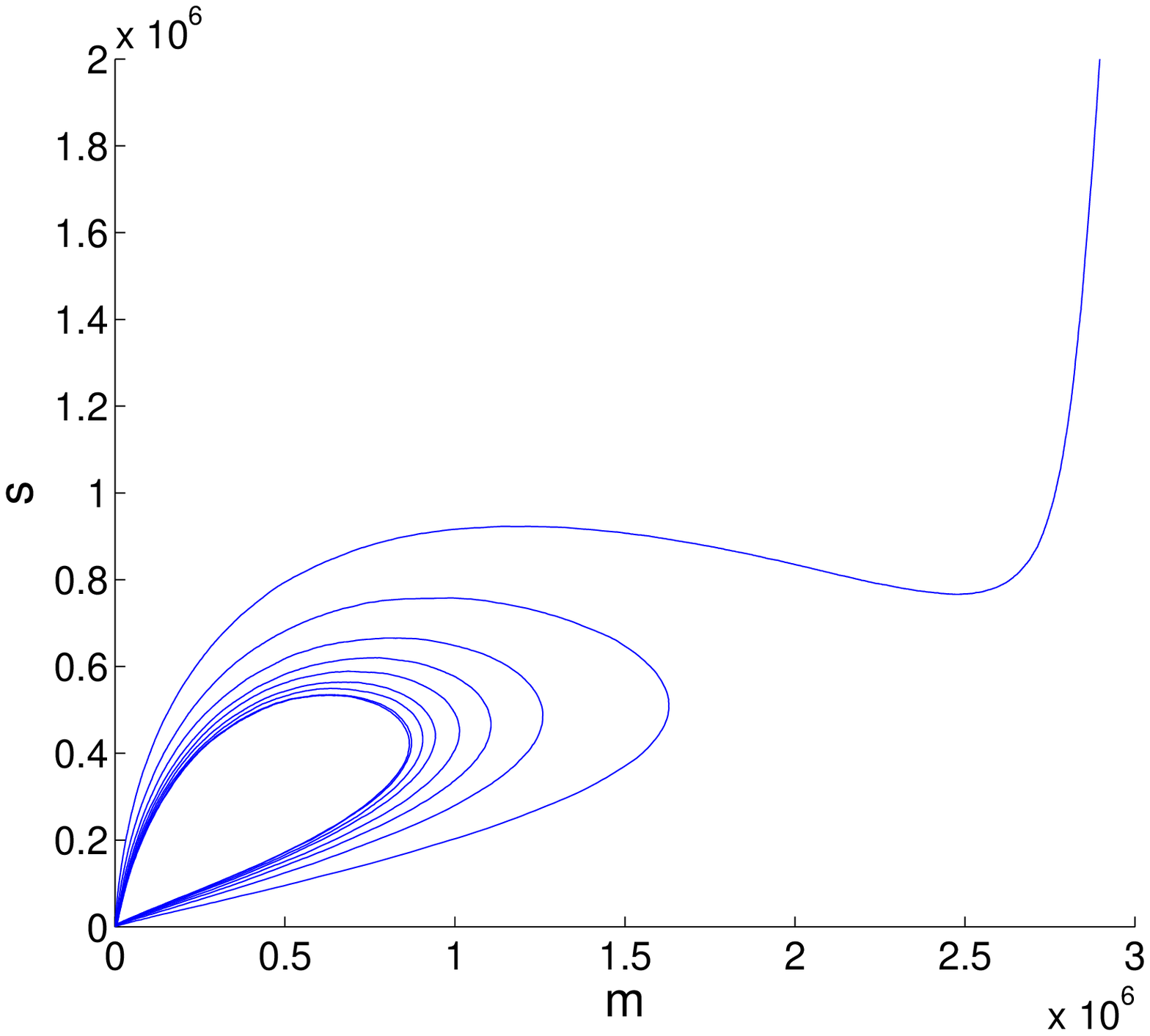}\newline
\caption{The projection of Fig. \ref{limit_cycle_g} onto the m-s plane.}
\label{limit_cycle_gms} \
\end{figure}

\begin{figure}[tbp]
\includegraphics[angle=-90,width=0.99\columnwidth]{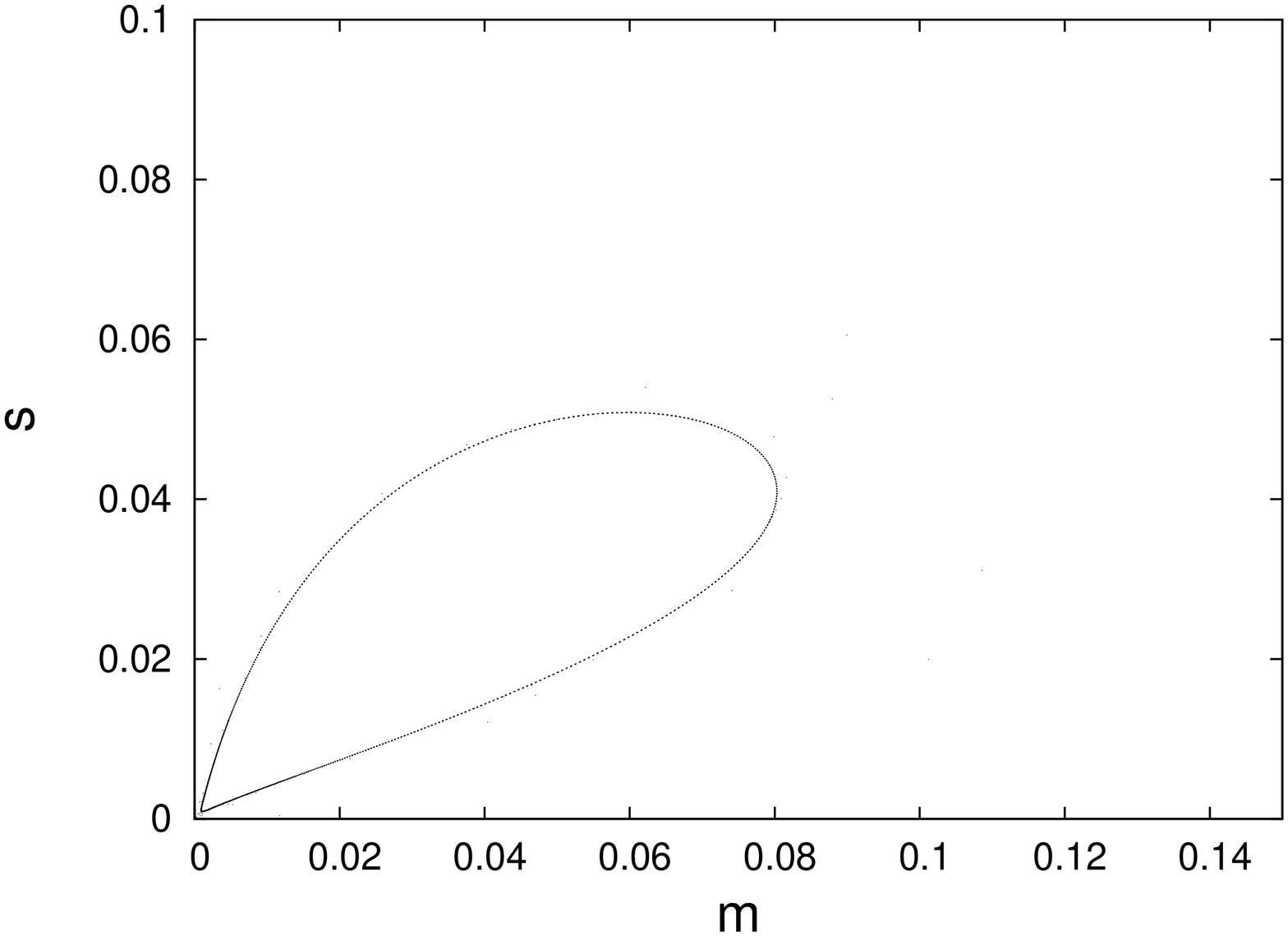}\newline
\caption{A limit cycle projected onto m-s plane in the mean field
theory with parameters $\phi_1 = 1, \phi_2 = 0.8, d_1 = 0.5, d_2 =%
0.9, d_3 = 0.03, a_1 = a_{21} = a_{31} = 0.1$, and $a_{22} = a_{32} =%
0.5.$} \label{limit_cycle_3ms} \
\end{figure}

\begin{figure}[tbp]
\includegraphics[width=0.99\columnwidth]{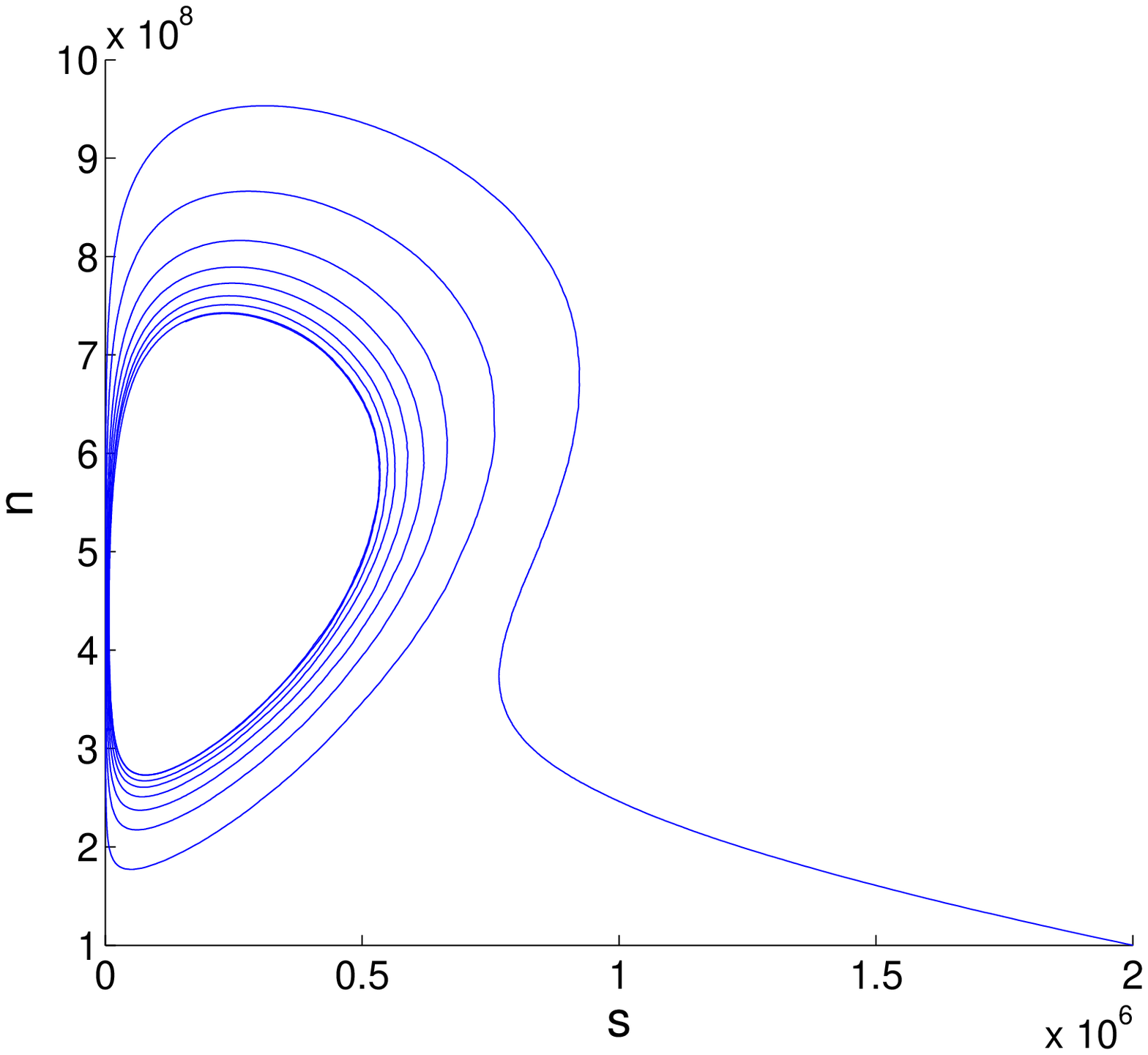}\newline
\caption{The projection of Fig. \ref{limit_cycle_g} onto the s-n plane.}
\label{limit_cycle_gsn} \
\end{figure}

\begin{figure}[tbp]
\includegraphics[angle=-90,width=0.99\columnwidth]{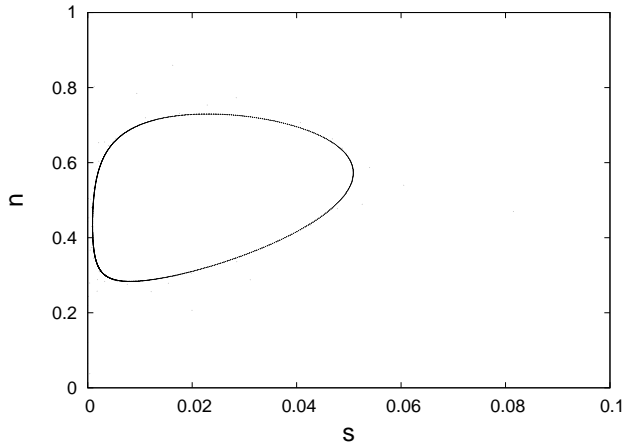}\newline
\caption{A limit cycle projected onto the s-n plane in the mean field
theory with parameters $\phi_1 = 1, \phi_2 = 0.8, d_1 = 0.5, d_2 =%
0.9, d_3 = 0.03, a_1 = a_{21} = a_{31} = 0.1$, and $a_{22} = a_{32} =%
0.5.$} \label{limit_cycle_3sn} \
\end{figure}

In Fig. \ref{limit_cycle_g}, we show a limit cycle observed in our
stochastic simulations. It is broadly consistent with the mean field
predictions, as can be noted easily by the obvious similarities between
Fig. \ref{limit_cycle_gms} and Fig. \ref{limit_cycle_3ms}, and Fig.
\ref{limit_cycle_gsn} and Fig. \ref{limit_cycle_3sn} (when we project
the three-dimensional phase space onto two dimensions),
whose relationship
is Eq. (\ref{mprime}), (\ref{sprime}) and (\ref{nprime}).
As expected, we notice fluctuations in the stochastic simulation. For
example, if Fig. \ref{limit_cycle_g} were shown in better resolution,
we could see that the curve wiggled around the limit cycle. Usually
fluctuation is two orders of magnitude smaller than the value it wiggles.
Hence we conclude that the limit cycle is inherent to the model and robust
to stochastic fluctuations, which serves to confirm the essential role of
lysogens in stabilizing the cycling in the populations.

\section{Parameters in the model}
\label{parameters}

Up to this point, all the parameters above or their values we have
explored are difficult to relate to experiment. The purpose of this
section is to bridge the gap.

The birth rate of the host $ b $ is medium-dependent. Usually the
expression of Lac proteins is highly suppressed by Lac repressors in a
lacose-free medium to optimize energy investment and metabolism of the
bacteria.  In the above two models, we have categorized the death of
the hosts to longevity and crowding. In fact, it is hard to mark a
watershed clearly. Instead, what is observed is a population-dependent
growth rate, which is a combined effect of $ b $, $ c $ and $ d $.
Herein, the rate $ d $ for the death of the host due to crowding is
introduced artificially to account for the actual population
dependence.  Thus, we are justified in assuming that the death rate of
the host due to longevity $ c $, which incorporates other physical and
non-density-dependent factors, is fixed with the variation in host
population. The growth rate for E. \textit{coli} may drop to $ 0.2 $ h$
^{-1} $ at $ 37^{\circ} $C when glycolate serves as the carbon source
but usually is in the range from $ 0.5 $ h$ ^{-1} $ to $ 2.0 $ h$ ^{-1}
$\cite{MARR91,HADA97}. The growth rate is species- and strain-specific,
which for \textit{Pseudoalteromonas} sp. strain SKA18 (accessible
number AF188330 in GenBank)\cite{MIDD00}, for example, is an order of
magnitude smaller. Similarly, lysis rate $ f $, lysogeny rate $ k $,
prophage induction rate $ q $, and replicate number per capita $ \beta
$, which are all under poor metabolism, are introduced manually to
characterize the population-dependent feature of the interactions in
order to leave their population-independent counterparts $ e $, $ h $,
$ p $ and $ \alpha $ fixed. In the case of virulent phages, such as one
in the family Siphoviridae\cite{MIDD01} attacking
\textit{Pseudoalteromonas} sp. strain SKA18\cite{MIDD00}, corresponding
to the lysis-only model, the reported lysis rate spans from $ 0.2 $ to $ 2.0 $ h$
^{-1} $ subject to the growth rate of the bacteria so that we can
estimate $ e $ to be on the order of $ 1.0 $ h$ ^{-1} $ and $ f $ to be
an order of magnitude smaller than $ e $.

For temperate phages in the lysogeny-lysis model, the spontaneous lysis
rate is far smaller, being of the order of $ 10^{-9} $ to $ 10^{-7} $
per generation per cell\cite{AURE02}. The percentage of lysogens is
assayed through prophage induction by the addition of mitomycin C, UV
radiation or other environmental conditions that may inhibit lambda
phage repressors. Under good metabolism the lysogeny rate $ h $ for $
\lambda $ phage infecting E. \textit{coli} and prophage induction rate
are on the order of $ 1 $ h$ ^{-1} $ and $ 2 $ h$ ^{-1} $,
respectively\cite{KOUR71}. Their counterparts under poor metabolism are
estimated to be one or two orders of magnitude smaller. For instance,
the prophage induction rate for log-phase marine lysogens\cite{JIAN98}
is on the order of $ 0.03 $ h$ ^{-1} $. Replicate number per capita $
\alpha $ is about $ 100 $ for phage $ \lambda $\cite{PTAS04}, and may
be up to $ 600 $ for phage W-14\cite{KROP70}, while $ \beta $ is about
20 or 30 for both. The death of free phage is quite rare, which may
result from the cleavage by proteins and depends on physical conditions
such as temperature, humidity and pH values. C. D. Jepson and J. B.
March\cite{JEPS04} reported that phage $ \lambda $ is highly stable,
whose half life in suspension ranges from 2.3 days at $ 4.2^{\circ} $C
to 36 days at $ 20^{\circ} $C. Even if we take the half life be one
day, the corresponding death rate $ g $ is on the order of $ 10^{-6} $
per second and can be suppressed by cooling down. Actually, the loss of
free phage in nature, to a great extent, is through diffusion since
bacteria are more immobile due to their large particle size compared to
that of phages. In laboratory, the death rate can be manipulated
through continuous dilution and washing out, and a wide range of death
rates can be realized.

When all the parameters are tuned properly, the limit cycle in the
lysogeny-lysis model is observable in experiment.
We estimate the period of the limit cycle to be on the order of days.
Take Fig. \ref{limit_cycle_gms} as an example. A cycle there is composed of about
$ 10,000 $ computational steps, in other words $ 10,000 $ events, which
corresponds to about $ 120 $ [time unit] in the simulation. In Fig.
\ref{limit_cycle_gms} the birth rate is $ 0.4 $ [time unit]$ ^{-1} $,
while in the real world the life cycle of an E. \textit{coli} in good
laboratory conditions, for example, is about half an hour, which is $ 2 $ hour
$ ^{-1} $. Hence the cycle is $ 120 \times 0.4 / 2 = 24 $ hours, which is one day.

\section{Conclusion}

We have derived the mean field population dynamics for host-phage
communities both without and with lysogens. In the lysis-only model, we
successfully obtained a description similar to the starting point
assumed by Weitz and Dushoff\cite{WEIT08b}, and we found that the phase
diagram was modified only slightly to the difference in good and poor
metabolism. In the lysogeny-lysis model, we identified the asymptotic
states, which included not only coexistence and extinction fixed
points, but population cycling of all microbes, lysogens and phagess.
Our findings support the notion that lysogens act as a reservoir and
are in principle amenable to experimental verification.  We simulated
the stochastic process using the Gillespie algorithm and verified the
robustness of our results to fluctuations, and especially demonstrated
the stability of the limit cycle.

Although complicated, our model inevitably makes some drastic
assumptions, in addition to the most severe of all---the omission of
spatial structure.  In particular, we treat  ``healthy" hosts and
lysogens in the same way regarding their natural birth, death and
crowding effect. However, experimentally, the expression of prophage
genes and the control of host gene expression by viral genes seem to
impart to lysogens econominization in their metabolism\cite{PAUL08}.
When unnecessary metabolic activities are suppressed, lysogens optimize
their energy expenses and therefore gain some survival advantage
compared to ``healthy" hosts in unfavorable conditions, which suggests
that the natural birth, death and crowding effects of lysogens are
distinct from those of ``healthy" hosts. Hence our model is a minimal
model that can capture the non-trivial role of lysogens in the population
dynamics of microbe-phage communities, in addition to the usual
predator-prey interactions.

This work can be extended in several ways, but perhaps the most
interesting are those which relate to the evolution of the field of
genes distributed amongst the microbes, viruses and lysogens.  Lysogens
are genome carriers of not only microbes but also prophages, capable of
yielding virus bursts when triggered by environmental stress. In this
way, the role of lysogens and viruses as a reservoir of genes is
mediated through phage infection and the lysogeny-lysis switch by the
metabolism of the host. The metabolism of the host is, in turn, to a
great extent influenced by environmental conditions.  Thus, this model
is a starting point for ecology-mediated evolution.  It is also useful
to stress that each individual microbe or virus constitutes a part of
another organism's environment.  Thus, the effects which our work
begins to treat, represent a microcosm of the intricate interplay
between ecology and evolution in microbe-virus communities.

\begin{acknowledgments}
We greatly thank Carl Woese, Ido Golding, Rachel Whitaker, Nicholas Chia, David
Reynolds, Nicholas Guttenberg, Patricio Jeraldo, Tom Butler and Maksim
Sipos for helpful discussions.  This work was supported in part by the
National Science Foundation through grant number NSF-0526747.
\end{acknowledgments}

\bibliographystyle{apsrev}
\bibliography{host_phage_bib}

\appendix
\section{Transition matrices for the lysis-only model}
\label{appA}

Here we provide the transition matrices, which are the probabilities for
the change in the population in each time step in the lysis-only model.

\begin{widetext}
\begin{equation}
T\left( m+1,n|m,n\right) =b\mu \left( 1-\nu \right) \left( 1-\omega
\right)
\dfrac{2m\left( K-m\right) }{K\left( K-1\right) }=\widetilde{b}m\left( 1-%
\dfrac{m}{K}\right) ;
\end{equation}

\begin{equation}
\widetilde{b}=\dfrac{2b\mu \left( 1-\nu \right) \left( 1-\omega \right) }{K-1%
}\approx \dfrac{2b\mu \left( 1-\nu \right) \left( 1-\omega \right)
}{K};
\end{equation}

\begin{equation}
T\left( m-1,n|m,n\right) =c\mu \left( 1-\nu \right) \omega
\dfrac{m}{K}+d\mu
\left( 1-\nu \right) \left( 1-\omega \right) \dfrac{m\left( m-1\right) }{%
K\left( K-1\right) }=\widetilde{c}m+\widetilde{d}m\left( \dfrac{m}{K}-\dfrac{%
1}{K}\right) \approx \widetilde{c}m+\widetilde{d}\dfrac{m^{2}}{K};
\end{equation}

\begin{equation}
\widetilde{c}=\dfrac{c\mu \left( 1-\nu \right) \omega }{K};
\end{equation}

\begin{equation}
\widetilde{d}=\dfrac{d\mu \left( 1-\nu \right) \left( 1-\omega \right) }{K-1}%
\approx \dfrac{d\mu \left( 1-\nu \right) \left( 1-\omega \right)
}{K};
\end{equation}

\begin{equation}
T\left( m-1,n+\alpha -1|m,n\right) =e\mu \nu \left( 1-\omega \right) \dfrac{%
2m\left( K-m\right) }{K\left( K-1\right)
}\dfrac{n}{W}=\widetilde{e}mn\left( 1-\dfrac{m}{K}\right) ;
\end{equation}

\begin{equation}
\widetilde{e}=\dfrac{2e\mu \nu \left( 1-\omega \right) }{\left( K-1\right) W}%
\approx \dfrac{2e\mu \nu \left( 1-\omega \right) }{KW}.
\label{etilde1}
\end{equation}

\begin{equation}
T\left( m-1,n+\beta -1|m,n\right) =f\mu \nu \left( 1-\omega \right) \dfrac{%
m\left( m-1\right) }{K\left( K-1\right) }\dfrac{n}{W}=\widetilde{f}\dfrac{%
m^{2}n}{K};
\end{equation}

\begin{equation}
\widetilde{f}=\dfrac{f\mu \nu \left( 1-\omega \right) }{\left( K-1\right) W}%
\approx \dfrac{f\mu \nu \left( 1-\omega \right) }{KW}.
\label{ftilde1}
\end{equation}

\begin{equation}
T\left( m,n-1|m,n\right) =g\left( 1-\mu \right) \nu \dfrac{n}{W}=\widetilde{g%
}n;
\end{equation}

\begin{equation}
\widetilde{g}=\dfrac{\left( 1-\mu \right) \nu }{W}.
\end{equation}
\end{widetext}

All the other transition matrixes are zero. Noting that all the events in
Table \ref{tab:lysis_events} are Markov processes, we know that the time
evolution for the probability with $m$ hosts and $n$ phages at time $t$ will
be

\begin{widetext}
\begin{align}
&\dfrac{d}{dt}P\left( m,n,t\right) =T\left( m,n|m-1,n\right)
P\left(
m-1,n,t\right) +T\left( m,n|m+1,n\right) P\left( m+1,n,t\right)\nonumber  \\
&+T\left( m,n|m+1,n+\alpha -1\right) P\left( m+1,n+\alpha
-1,t\right)
+T\left( m,n|m+1,n+\beta -1\right) P\left( m+1,n+\beta -1,t\right)\nonumber  \\
&+T\left( m,n|m,n+1\right) P\left( m,n+1,t\right) -[T\left(
m+1,n|m,n\right) +T\left( m-1,n|m,n\right) +T\left( m-1,n+\alpha
-1|m,n\right)\nonumber  \\
&+T\left( m-1,n+\beta -1|m,n\right) +T\left( m,n-1|m,n\right)
]P\left( m-1,n,t\right) .
\end{align}
\end{widetext}%

Applying summations according to Eq. (\ref{mn}), we will get

\begin{subequations}
\allowdisplaybreaks
\begin{align}
\dfrac{d\left\langle m\right\rangle }{dt}&= \left\langle T\left(
m+1,n|m,n\right) \right\rangle -\left\langle T\left( m-1,n|m,n\right)
\right\rangle  \notag \\
&\qquad -\left\langle T\left( m-1,n+\alpha -1|m,n\right) \right\rangle  \notag \\
&\qquad -\left\langle T\left( m-1,n+\beta -1|m,n\right) \right\rangle  \notag \\
&\approx \left( \widetilde{b}+\widetilde{d}\right) \left\langle
m\right\rangle \left( 1-\dfrac{\left\langle m\right\rangle }{K}\right)
-\left( \widetilde{c}+\widetilde{d}\right) \left\langle m\right\rangle
\notag \\
&\qquad -\widetilde{e}\left\langle m\right\rangle \left\langle n\right\rangle %
\left[ 1-\left( 1-\dfrac{\widetilde{f}}{\widetilde{e}}\right) \dfrac{%
\left\langle m\right\rangle }{K}\right]; \\
\dfrac{d\left\langle n\right\rangle }{dt}&=\left( \alpha -1\right)
\left\langle T\left( m-1,n+\alpha -1|m,n\right) \right\rangle  \notag \\
&\qquad + \left( \beta -1\right) \left\langle T\left( m-1,n+\beta -1|m,n\right)
\right\rangle  \notag \\
&\qquad - \left\langle T\left( m,n-1|m,n\right) \right\rangle  \notag \\
&= \left( \alpha -1\right) \ \widetilde{e}\left\langle m\right\rangle
\left\langle n\right\rangle \left[ 1-\left( 1-\dfrac{\left( \beta -1\right)
\widetilde{f}}{\left( \alpha -1\right) \widetilde{e}}\right) \dfrac{%
\left\langle m\right\rangle }{K}\right]  \notag \\
&\qquad - \widetilde{g}\left\langle n\right\rangle .
\end{align}
\end{subequations}%

Let

\begin{subequations}
\allowdisplaybreaks
\begin{align}
r &= \widetilde{b}+\widetilde{d}; \\
\phi &= \widetilde{e}; \\
\gamma &= \alpha -1; \\
d_{m} &= \widetilde{c}+\widetilde{d}; \\
d_{n} &= \widetilde{g}; \\
a_{m} &= 1-\dfrac{\widetilde{f}}{\widetilde{e}}; \\
a_{n} &= 1-\dfrac{\left( \beta -1\right) \widetilde{f}}{\left( \alpha
-1\right) \widetilde{e}};
\end{align}
\end{subequations}%

\noindent
which is Eq. (\ref{parameter1}), we can arrive at Eq. (\ref{lysis1}).

\section{Transition matrices for the lysogeny-lysis model}
\label{appB}

In this appendix, we provide details for the derivations of the lysogeny-lysis
model.

According to Table \ref{tab:lysogeny_events} and Table \ref%
{tab:lysogeny_prob}, we can obtain the following non-zero transition
matrixes:

\begin{widetext}
\begin{equation}
T\left( m+1,s,n|m,s,n\right) =b\mu \left( 1-\nu \right) \left(
1-\omega
\right) \dfrac{2m\left( K-m-s\right) }{K\left( K-1\right) }=\widetilde{b}%
m\left( 1-\dfrac{m+s}{K}\right) ;
\end{equation}

\begin{equation}
\widetilde{b}=\dfrac{2b\mu \left( 1-\nu \right) \left( 1-\omega \right) }{K-1%
}\approx \dfrac{2b\mu \left( 1-\nu \right) \left( 1-\omega \right)
}{K}.
\end{equation}

\begin{equation}
T\left( m,s+1,n|m,s,n\right) =b\mu \left( 1-\nu \right) \left(
1-\omega
\right) \dfrac{2s\left( K-m-s\right) }{K\left( K-1\right) }=\widetilde{b}%
s\left( 1-\dfrac{m+s}{K}\right) ;
\end{equation}

\begin{align}
T\left( m-1,s,n|m,s,n\right) &=c\mu \left( 1-\nu \right) \omega \dfrac{m}{K}%
+d\mu \left( 1-\nu \right) \left( 1-\omega \right) \dfrac{m\left(
m-1\right) }{K\left( K-1\right) }+\dfrac{1}{2}d\mu \left( 1-\nu
\right) \left( 1-\omega
\right) \dfrac{2ms}{K\left( K-1\right) } \nonumber\\
&=\widetilde{c}m+\widetilde{d}m\dfrac{m+s}{K}.
\end{align}

\begin{equation}
\widetilde{c}=\dfrac{c\mu \left( 1-\nu \right) \omega }{K};
\end{equation}

\begin{equation}
\widetilde{d}=\dfrac{d\mu \left( 1-\nu \right) \left( 1-\omega \right) }{K-1}%
\approx \dfrac{d\mu \left( 1-\nu \right) \left( 1-\omega \right)
}{K}.
\end{equation}

\begin{align}
T\left( m,s-1,n|m,s,n\right) &=c\mu \left( 1-\nu \right) \omega \dfrac{s}{K}%
+d\mu \left( 1-\nu \right) \left( 1-\omega \right) \dfrac{s\left(
s-1\right) }{K\left( K-1\right) }+\dfrac{1}{2}d\mu \left( 1-\nu
\right) \left( 1-\omega
\right) \dfrac{2ms}{K\left( K-1\right) } \nonumber\\
&=\widetilde{c}s+\widetilde{d}s\dfrac{m+s}{K}.
\end{align}

\begin{equation}
T\left( m-1,s,n+\alpha +1|m,s,n\right) =e\mu \nu \left( 1-\omega
\right)
\dfrac{2m\left( K-m-s\right) }{K\left( K-1\right) }\dfrac{n}{W}=\widetilde{e}%
mn\left( 1-\dfrac{m+s}{K}\right) ;
\end{equation}

\begin{equation}
\widetilde{e}=\dfrac{2e\mu \nu \left( 1-\omega \right) }{\left( K-1\right) W}%
\approx \dfrac{2e\mu \nu \left( 1-\omega \right) }{KW}.
\end{equation}

\begin{equation}
T\left( m-1,s,n+\beta +1|m,s,n\right) =f\mu \nu \left( 1-\omega
\right) \dfrac{m\left( m-1\right) }{K\left( K-1\right)
}\dfrac{n}{W}+f\mu \nu \left(
1-\omega \right) \dfrac{2ms}{K\left( K-1\right) }\dfrac{n}{W}=\widetilde{f}mn%
\dfrac{m+2s}{K};
\end{equation}

\begin{equation}
\widetilde{f}=\dfrac{f\mu \nu \left( 1-\omega \right) }{\left( K-1\right) W}%
\approx \dfrac{f\mu \nu \left( 1-\omega \right) }{KW}.
\end{equation}

\begin{align}
&T\left( m-1,s+1,n-1|m,s,n\right) =h\mu \nu \left( 1-\omega \right) \dfrac{%
2m\left( K-m-s\right) }{K\left( K-1\right) }\dfrac{n}{W}+k\mu \nu
\left( 1-\omega \right) \dfrac{m\left( m-1\right) }{K\left(
K-1\right) }\dfrac{n}{W}\nonumber\\
&+k\mu \nu \left( 1-\omega \right) \dfrac{2ms}{K\left( K-1\right) }\dfrac{n}{W%
}=\widetilde{h}mn\left( 1-\dfrac{m+s}{K}\right) +\widetilde{k}mn\dfrac{m+2s%
}{K};
\end{align}

\begin{equation}
\widetilde{h}=\dfrac{2h\mu \nu \left( 1-\omega \right) }{\left( K-1\right) W}%
\approx \dfrac{2h\mu \nu \left( 1-\omega \right) }{KW};
\end{equation}

\begin{equation}
\widetilde{k}=\dfrac{k\mu \nu \left( 1-\omega \right) }{\left( K-1\right) W}%
\approx \dfrac{k\mu \nu \left( 1-\omega \right) }{KW}.
\end{equation}

\begin{equation}
T\left( m,s-1,n+\alpha |m,s,n\right) =p\mu \left( 1-\nu \right)
\left(
1-\omega \right) \dfrac{2s\left( K-m-s\right) }{K\left( K-1\right) }=%
\widetilde{p}s\left( 1-\dfrac{m+s}{K}\right) ;
\end{equation}

\begin{equation}
\widetilde{p}=\dfrac{2p\mu \left( 1-\nu \right) \left( 1-\omega \right) }{K-1%
}\approx \dfrac{2p\mu \left( 1-\nu \right) \left( 1-\omega \right)
}{K}.
\end{equation}

\begin{equation}
T\left( m,s-1,n+\beta |m,s,n\right) =q\mu \left( 1-\nu \right)
\left( 1-\omega \right) \dfrac{s\left( s-1\right) }{K\left(
K-1\right) }+q\mu
\left( 1-\nu \right) \left( 1-\omega \right) \dfrac{2ms}{K\left( K-1\right) }%
=\widetilde{q}s\dfrac{2m+s}{K};
\end{equation}

\begin{equation}
\widetilde{q}=\dfrac{q\mu \left( 1-\nu \right) \left( 1-\omega \right) }{K-1}%
\approx \dfrac{q\mu \left( 1-\nu \right) \left( 1-\omega \right)
}{K}.
\end{equation}

\begin{equation}
T\left( m,s,n-1|m,s,n\right) =g\left( 1-\mu \right) \nu \dfrac{n}{W}=%
\widetilde{g}n;
\end{equation}

\begin{equation}
\widetilde{g}=\dfrac{\left( 1-\mu \right) \nu }{W}.
\end{equation}
\end{widetext}

Ignoring fluctuations and correlations, we derive the populations
dynamics at the mean field level. The time evolution for population size is

\begin{widetext}
\begin{subequations}
\allowdisplaybreaks
\begin{align}
\dfrac{d\left\langle m\right\rangle }{dt}&=\left\langle T\left(
m+1,s,n|m,s,n\right) \right\rangle -\left\langle T\left(
m-1,s,n|m,s,n\right) \right\rangle -\left\langle T\left(
m-1,s,n+\alpha
-1|m,s,n\right) \right\rangle  \nonumber\\
&\qquad -\left\langle T\left( m-1,s,n+\beta -1|m,s,n\right) \right\rangle
-\left\langle T\left( m-1,s+1,n-1|m,s,n\right) \right\rangle \nonumber \\
&=\left( \widetilde{b}+\widetilde{d}\right) \left\langle
m\right\rangle
\left( 1-\dfrac{\left\langle m\right\rangle +\left\langle s\right\rangle }{K}%
\right) -\left( \widetilde{c}+\widetilde{d}\right) \left\langle
m\right\rangle\nonumber\\
&\qquad -\left( \widetilde{e}+\widetilde{h}\right) \left\langle
m\right\rangle \left\langle n\right\rangle \left\{1-\dfrac{1}{K}\left[\left( 1-\dfrac{%
\widetilde{f}+\widetilde{k}}{\widetilde{e}+\widetilde{h}}\right)
\left\langle m\right\rangle +\left( 1-2\cdot \dfrac{\widetilde{f}+\widetilde{%
k}}{\widetilde{e}+\widetilde{h}}\right) \left\langle s\right\rangle
\right]\right\}; \\
\dfrac{d\left\langle s\right\rangle }{dt}&=\left\langle T\left(
m,s+1,n|m,s,n\right) \right\rangle -\left\langle T\left(
m,s-1,n|m,s,n\right) \right\rangle +\left\langle T\left(
m-1,s+1,n-1|m,s,n\right) \right\rangle  \nonumber\\
&\qquad -\left\langle T\left( m,s-1,n+\alpha |m,s,n\right) \right\rangle
-\left\langle T\left( m,s-1,n+\beta |m,s,n\right) \right\rangle  \nonumber\\
&=\left( \widetilde{b}+\widetilde{d}\right) \left\langle
s\right\rangle
\left( 1-\dfrac{\left\langle m\right\rangle +\left\langle s\right\rangle }{K}%
\right)  -\left( \widetilde{c}+\widetilde{d}\right) \left\langle
s\right\rangle \nonumber\\
&\qquad -\widetilde{h}\left\langle m\right\rangle \left\langle
n\right\rangle \left\{1-\dfrac{1}{K}\left[\left( 1-\dfrac{\widetilde{k}}{\widetilde{h}}%
\right) \left\langle m\right\rangle +\left( 1-2\cdot \dfrac{\widetilde{k}}{%
\widetilde{h}}\right) \left\langle s\right\rangle \right]\right\} \nonumber\\
&\qquad-\widetilde{p}\left\langle s\right\rangle \left\{1-\dfrac{1}{K}\left[\left(
1-2\cdot \dfrac{\widetilde{q}}{\widetilde{p}}\right) \left\langle
m\right\rangle +\left( 1-\dfrac{\widetilde{q}}{\widetilde{p}}\right)
\left\langle s\right\rangle \right]\right\}; \\
\dfrac{d\left\langle n\right\rangle }{dt}&=\left( \alpha -1\right)
\left\langle T\left( m-1,s,n+\alpha -1|m,s,n\right) \right\rangle
+\left( \beta -1\right) \left\langle T\left( m-1,s,n+\beta
-1|m,s,n\right) \right\rangle\nonumber\\
&\qquad -\left\langle T\left( m-1,s+1,n-1|m,s,n\right) \right\rangle
+\alpha \left\langle T\left( m,s-1,n+\alpha |m,s,n\right)
\right\rangle +\beta \left\langle T\left( m,s-1,n+\beta
|m,s,n\right) \right\rangle\nonumber\\
&\qquad -\left\langle T\left( m,s,n-1|m,s,n\right) \right\rangle  \nonumber\\
&=\left[ \left( \alpha -1\right) \widetilde{e}-\widetilde{h}\right]
\left\langle m\right\rangle \left\langle n\right\rangle
\left\{1-\dfrac{1}{K}
\left[ \left( 1-\dfrac{\left( \beta -1\right) \widetilde{f}-%
\widetilde{k}}{\left( \alpha -1\right)
\widetilde{e}-\widetilde{h}}\right) \left\langle m\right\rangle
+\left( 1-2\cdot \dfrac{\left( \beta -1\right)
\widetilde{f}-\widetilde{k}}{\left( \alpha -1\right) \widetilde{e}-%
\widetilde{h}}\right) \left\langle s\right\rangle \right]\right\} \nonumber\\
&\qquad +\alpha \widetilde{p}\left\langle s\right\rangle \left \{1-\dfrac{1}{K}%
\left[\left( 1-2\cdot \dfrac{\beta \widetilde{q}}{\alpha
\widetilde{p}}\right) \left\langle m\right\rangle +\left(
1-\dfrac{\beta \widetilde{q}}{\alpha
\widetilde{p}}\right) \left\langle s\right\rangle \right]\right\}-\widetilde{g}%
\left\langle n\right\rangle .
\end{align}
\label{lysogeny2}
\end{subequations}
\end{widetext}%

Let
\begin{subequations}
\allowdisplaybreaks
\begin{align}
r &=\widetilde{b}+\widetilde{d}; \\
d_{1} &=\widetilde{c}+\widetilde{d}; \\
d_{2} &=\widetilde{p}; \\
d_{3} &=\widetilde{g}; \\
\phi _{1} &=\widetilde{e}+\widetilde{h}; \\
\phi _{2} &=\widetilde{h}; \\
a_{1} &=\dfrac{\widetilde{f}+\widetilde{k}}{\widetilde{e}+\widetilde{h}}; \\
a_{21} &=\dfrac{\widetilde{k}}{\widetilde{h}}; \\
a_{22} &=\dfrac{\widetilde{q}}{\widetilde{p}}; \\
a_{31} &=\dfrac{\left( \beta -1\right) \widetilde{f}-\widetilde{k}}{\left(
\alpha -1\right) \widetilde{e}-\widetilde{h}}; \\
a_{32} &=\dfrac{\beta \widetilde{q}}{\alpha \widetilde{p}};
\end{align}%
\end{subequations}%

\noindent
which is Eq. (\ref{parameter2}), and omit angle-brackets for simplicity, Eq. (\ref{lysogeny2}) can be written as Eq. (\ref{lysogeny1}).

\end{document}